\documentclass[lettersize,journal]{IEEEtran}
\usepackage{amsmath,amsfonts}
\usepackage{algorithmic}
\usepackage{algorithm}
\usepackage{multirow}
\usepackage{array}
\usepackage{graphicx}
\usepackage{subcaption}
\usepackage{textcomp}
\usepackage{stfloats}
\usepackage{url}
\usepackage{verbatim}

\usepackage{cite}
\usepackage{pifont}

\usepackage{color}

\usepackage{xcolor}
\usepackage[
    colorlinks=true,
    allcolors=blue,
    breaklinks=true 
]{hyperref}

\begin{document}

\title{Threat-Aware Energy-Efficient Deployment for Dynamic UAV Networks: A Multi-Agent RL Approach}

\author{
 	Faisal~Al-Kamali, Hussein~A.~Ammar,~\IEEEmembership{Member,~IEEE}, Francois ~Chan,~\IEEEmembership{Senior Member,~IEEE}, James~H.~Bayes, \IEEEmembership{Member,~IEEE}, Yasser Gadallah, \IEEEmembership{Senior Member,~IEEE}, and Mohamed H. Ahmed, \IEEEmembership{Senior Member,~IEEE}
    \vspace{-2.5em} 

 \thanks{Manuscript received xxxx, xxxx.}  
  \thanks{Faisal Al-Kamali, Francois Chan, and Mohamed H. Ahmed are with the School of Electrical Engineering and Computer Science, University of Ottawa, Ottawa, ON K1N 6N5, Canada (e-mails: falkamal@uottawa.ca, chan-f@rmc.ca, and  mahme3@uottawa.ca).}%
  \thanks{Faisal Al-Kamali is also with the Department of Electrical Engineering, Ibb University, Ibb, Yemen.}
  \thanks{Francois~Chan, Hussein~A.~Ammar, and James~H.~Bayes  are with the Department of Electrical and Computer Engineering, Royal Military College, Kingston, ON, K7K 7B4, Canada (e-mails: chan-f@rmc.ca, hussein.ammar@rmc-cmr.ca, james.bayes@rmc-cmr.ca}
  \thanks{Yasser Gadallah is with the Department of Electronics and Communications Engineering, The American University in Cairo, New Cairo 11835, Egypt (email: ygadallah@ieee.org).}
}





\maketitle

\begin{abstract}
Ensuring operational safety in threat-prone environments remains a critical challenge for multi-UAV networks serving as aerial base stations. This paper proposes an efficient framework to maximize global energy efficiency (EE) while promoting safe operation through threat-aware clustering and reward-based safety enforcement. The proposed framework is executed in three steps. First, a threat-aware K-means (TAKM) algorithm determines the minimum required UAVs and computes safe initial placements. Second, an optimal matching stage assigns physical UAVs to these centroids to minimize energy expenditure. Third, a threat-aware multi-agent twin delayed deep deterministic policy gradient (MATD3) algorithm dynamically optimizes trajectories, power, and user associations. Simulation results show that the proposed framework achieves zero observed safety violations in the considered scenarios while achieving superior EE and faster convergence than other learning methods and non-clustering baselines. Compared to heuristic optimization, the proposed framework outperforms the greedy particle swarm optimization (GPSO) and achieves performance comparable to that of the optimized PSO (OPSO), while incurring significantly lower online deployment computational complexity. Furthermore, the proposed framework demonstrates effective generalization to unseen user distributions, large UAV fleets, and different threat geometries, while maintaining zero safety violations.

\end{abstract}

\begin{IEEEkeywords}
Unmanned aerial vehicle (UAV), trajectory design, resource allocation, threat-aware clustering, multi-agent reinforcement learning.
\end{IEEEkeywords}

\section{Introduction}

\subsection{Background}
\IEEEPARstart{U}{nmanned} aerial vehicles (UAVs), commonly known as drones, have emerged as an important technology for next-generation wireless communication networks. Owing to their mobility, flexible deployment, and relatively low operational cost, UAVs can provide rapid and adaptive connectivity in scenarios where terrestrial infrastructure is unavailable, damaged, or impractical \cite{8579209, 11362961}. Their ability to establish favorable line-of-sight (LoS) links with ground users (GUs) reduces propagation loss and transmit power requirements, thereby enabling reliable and energy-efficient communication. These advantages make UAV-assisted systems a promising solution for diverse applications, including disaster recovery, emergency response, rural connectivity, and temporary hotspot provisioning \cite{7470933}.

UAVs can be used as temporary aerial base stations, such as in major events or emergencies, to support wireless communication, since they can be flexibly deployed and repositioned to provide on-demand connectivity for GUs. Consequently, many works have focused on single-UAV deployment and placement optimization in two- or three-dimensional space \cite{7918510, 6863654}. These studies have shown clear advantages, such as improved coverage, reliable LoS links, and enhanced network performance. UAVs can also play a vital role in enhancing security within threat-prone environments. In  \cite{11329043}, we present a novel UAV placement strategy that improves secure wireless communication by enforcing spatial threat exclusion.  Nevertheless, single-UAV systems remain constrained by their limited communication, caching, and computational capabilities, which restrict their ability to support mission-critical services or serve a large number of GUs \cite{9767553}.

To overcome these limitations, the deployment of multi-UAV communication systems has been proposed \cite{8247211, 8676325}. In such systems, multiple UAVs cooperatively serve GUs distributed over a large area, enabling simultaneous service. This cooperative approach not only increases system throughput but also reduces latency and improves scalability, thereby addressing challenges that single-UAV systems cannot overcome. However, the deployment of multi-UAV communication system entails several technical challenges, including optimal placement, trajectory design, and energy management \cite{8038869, 8663615}. In particular, the limited on-board battery capacity significantly restricts their ability to provide sustained coverage.

\subsection{Motivation and Novelty}

Ensuring the survivability and operational integrity of UAVs while maintaining reliable connectivity for GUs constitutes the primary motivation of this work. Specifically, we aim to deploy a multi-UAV network capable of providing reliable communication services in environments characterized by stringent safety constraints and the presence of threat zones. To address this challenge, the problem is formulated as a joint optimization of UAV trajectory design, transmit power allocation, and dynamic user association. Although this problem has attracted increasing research attention  \cite{8501974}, \cite{ 8618602, 8068199}, most existing studies do not incorporate threat avoidance as a fundamental design constraint, thereby limiting their applicability in safety-critical scenarios.

\begin{table*}[t]
	\centering
	\caption{Comparison Between the Proposed Framework and Related Works}
	\label{tab:comparison_all}
	\setlength{\tabcolsep}{5pt}
	\renewcommand{\arraystretch}{1.2}
	\begin{tabular}{|l|c|c|c|c|c|c|}
		\hline
		\textbf{Aspect} & \textbf{\cite{9767553}} & \textbf{\cite{9826431}} & \textbf{\cite{9209079}} & \textbf{\cite{9152044}} & \textbf{\cite{8727504}} & \textbf{This Work} \\
		\hline
		User distribution & Uniform & Uniform & Uniform & Uniform  & Real-world  & Non-uniform \\
		\hline
		Threat-zone awareness & \ding{55} & \ding{55} & \ding{55} & \ding{55} & \ding{55} & \ding{51} \\
		\hline
		UAV count optimization & \ding{55} & \ding{55} & \ding{55} & \ding{55} & \ding{55} & \ding{51} \\
		\hline
		User association strategy & Dynamic & Fixed & Dynamic & Fixed  & Fixed & Dynamic \\
		\hline
		Clustering scheme & None & MEM & None & Hotspot-based & GAK-means & TAKM \\
		\hline
		Operational area & $1000 \times 1000$ m$^2$ & $1000 \times 1000$ m$^2$ & $100 \times 100$ m$^2$ & $500 \times 500$ m$^2$ & $2000 \times 2000$ m$^2$ & $4000 \times 4000$ m$^2$ \\
		\hline
	\end{tabular}
	\vspace{-1em}
\end{table*}

Conventional optimization and iterative methods  \cite{8678650},  \cite{8422581, 9222571} focus strictly on maximizing coverage and connectivity while ignoring the physical safety of the UAVs. Although, heuristic methods like particle swarm optimization (PSO) are more efficient for joint trajectory and power optimization, they are also unsuitable for real-time deployment. Their re-optimization-based nature requires iterative updates from scratch for every environmental change, incurring latency that exceeds the network's coherence time in high-mobility scenarios. Reinforcement learning (RL) has recently emerged as an effective approach in multi-UAV networks, capable of learning and adapting from experience without requiring complete knowledge of the network environment  \cite{9767553}, \cite{9826431,8676325}.

Despite these advantages, existing RL approaches for UAV trajectory and resource optimization often neglect critical safety considerations, such as threat-zone restrictions, which are essential for practical deployment in risk-sensitive environments. As a result, these approaches are not suitable for scenarios that involve threat zones such as wildfire regions or hazardous areas. Moreover,  conventional K-Means (CKM) prioritize geometric distance alone, which often leads to deploying UAVs inside threat zones. This creates an immediate risk of mission failure that current RL models cannot fix, as they typically lack a safety-aware initialization phase.

To bridge this gap, we propose a three-stage  framework that integrates  threat avoidance into every phase. The main novelty lies in the threat-aware K-means (TAKM) algorithm, which replaces standard clustering with a projection-based approach that forces UAV centroids to remain within verified safe regions. Unlike previous methods, TAKM embeds threat avoidance directly into the optimization process. This safe deployment is then bridged via an optimal matching scheme to a multi-agent twin delayed deep deterministic policy gradient (MATD3) for dynamic control. The key novelty is the integration of threat-aware clustering, optimal deployment, and multi-agent RL (MARL) into a unified framework that enforces safety constraints throughout all phases of UAV network operation.

 \subsection{Related Works}
Recent research has extensively investigated UAV-assisted wireless communications with a focus on trajectory and power optimization.  The study in {\cite{8247211} considers a multi-UAV wireless communication system and proposes an iterative algorithm that jointly optimizes user scheduling, UAV trajectory, and power control to maximize the minimum downlink throughput. Beyond throughput optimization, several studies have focused on secure UAV communications by co-designing trajectories and transmit power. In\cite{8501974}, a four-node system with a mobile UAV relay is investigated, employing an alternating optimization framework to maximize secrecy rate. Similarly, Zhang et al. \cite{8618602} propose a joint trajectory and power control approach to improve secrecy performance, while the work in \cite{8678650} considers UAV-ground communications in the presence of no-fly zones and develops a low-complexity algorithm to enhance both secrecy rate and computational efficiency. In addition to security-focused works,  \cite{8068199} investigates a UAV relay network with amplify-and-forward operation and proposes a low-complexity joint trajectory and power control design to minimize outage probability. Furthermore, \cite{9222571} examines a UAV-assisted backscatter communication network, with simulations showing significant gains over other benchmark schemes. In \cite{8648498}, the authors jointly optimize UAV 3D trajectory, transmit power, and scheduling variables to improve system performance. 

However, conventional optimization-based methods have high computational complexity and require full knowledge of environmental models, which may not be available in practice. To overcome these limitations, RL-based approaches have been increasingly considered for scenarios with partial or no channel state information. RL can be applied in both single-agent  \cite{10409538} and multi-agent \cite{9826431} settings, enabling UAVs to learn optimal policies for trajectory, power control, and resource allocation without requiring full knowledge of the environment. Also, tabular Q-learning methods have been applied for UAV placement and energy-efficient user association, as in \cite{10409538},  \cite{9273081,9310353}. However, the discrete state-action representation of these methods limits their applicability to continuous control problems~\cite{11095795}, such as power allocation and trajectory design, and lead to high space/computational complexity~\cite{10462006}.

Recently, the actor-critic framework \cite{Heess2012} has been employed in UAV trajectory and resource optimization. In \cite{Lillicrap2015}, deep deterministic policy gradient (DDPG) with deterministic policies provides a stable solution, and its multi-agent extension (MADDPG) \cite{Lowe2017} enables coordinated optimization in multi-UAV scenarios. MADDPG has been effectively utilized for joint UAV trajectory design, user fairness, load balancing, and energy efficiency \cite{9209079}, as well as for secure multi-UAV communications, including combined trajectory and transmit power optimization under minimum quality of service (QoS) requirements \cite{9161257,9152044}. In \cite{9826431}, MATD3 has been introduced, where a clustering-aided approach is developed to jointly optimize UAV trajectory, and transmit power, achieving faster convergence and improved performance.

Recent advancements in DRL-aided UAV networks have explored various facets of system optimization. For instance, learning-based radio resource management was investigated in \cite{9475535}, while NOMA-assisted trajectory scheduling was explored in \cite{10349347}. Meanwhile, path planning and trajectory control frameworks for MEC-integrated UAVs were proposed in \cite{11301845}. More recently, the focus has shifted toward distributed many-agent perspectives to handle scalability and safety constraints in large-scale swarms \cite{10552119}. Hybrid discrete-continuous MARL frameworks for UAV swarms were investigated in \cite{11391547}, where joint trajectory control, power allocation, and communication resource management were optimized using hybrid action representations.

Unlike existing DRL- and MARL-based methods that assume threat-free settings, the key novelty of this work is its end-to-end energy-efficient approach for UAV network optimization in threat-prone environments. This approach enforces safety constraints throughout all phases of UAV network operation. Specifically, we integrate the TAKM algorithm, optimal matching of UAVs to safe centroids, and threat-aware MATD3 into a unified framework.

As highlighted in Table~\ref{tab:comparison_all}, which compares our framework with existing works \cite{9767553,9826431}, \cite{9209079,9152044}, \cite{8727504}, our proposed approach introduces several key innovations. While \cite{9826431} presented a clustering-aided MARL method, it assumes uniformly distributed users and does not consider threat zones. In contrast, our framework is designed for non-uniform user distributions and is the only one that incorporates threat-zone awareness. Furthermore, our work addresses the optimization of the number of UAVs, a practical necessity absent in other referenced studies. Unlike the static user associations used in \cite{9826431,9152044}, our method supports dynamic association updates at each time step based on system conditions. Finally, our work replaces conventional clustering schemes with the safety-critical TAKM algorithm, ensuring safety during both initialization and iterative updates. These enhancements, summarized in  Table~\ref{tab:comparison_all}, result in a practical and safety-aware UAV deployment framework suitable for real-world constrained environments.

\subsection{Contribution}
The primary contribution of this paper is a risk-aware RL-based UAV placement framework that adapts to dynamic user association while maintaining low online complexity. The contributions of this paper are as follows:
\begin{enumerate}

     \item We formulate an energy-efficiency maximization problem for UAV-assisted networks under explicit safety and QoS constraints, including minimum rate requirements, inter-UAV collision avoidance, and threat-zone avoidance. The safety and QoS requirements are formulated as hard constraints in the underlying optimization problem. However, MATD3 handles safety through a reward-based penalty rather than a hard action-space constraint. TAKM provides safe initialization and centroid updates, while the safety penalty discourages trajectory violations during learning.

    \item We propose a TAKM clustering algorithm that embeds safety constraints directly into the clustering process, ensuring that all intermediate and final UAV placement points remain outside unsafe regions. TAKM is further integrated with a constraint-aware procedure to determine the minimum feasible number of UAVs required to satisfy coverage and QoS constraints.

    \item We develop a hybrid deployment and control framework that combines TAKM-based safe pre-placement with a MATD3-based learning approach, where UAV trajectories and transmit powers are optimized through continuous-action learning and user associations are deterministically updated.
\end{enumerate}

\subsection{Organization and Notation}
The rest of the paper is organized as follows. Section II presents the system model. Section III formulates the problem. Section IV introduces the clustering and UAV matching through an Algorithm. Section V describes the proposed threat-aware MARL learning process and presents the computational complexity analysis. Section VI provides the simulation settings and performance evaluation. Finally, Section VII concludes the paper and outlines future research directions. The list of notations is illustrated in \textcolor{blue}{Table~\ref{tab:symbols}}.

\begin{table*}[t]
\centering
\caption{List of Notations}
\label{tab:symbols}
\begin{tabular}{ll|ll|ll}
\hline
\textbf{Symbol} & \textbf{Description} &
\textbf{Symbol} & \textbf{Description} &
\textbf{Symbol} & \textbf{Description} \\
\hline
$a,b$ & Environment-dependent parameters &
$a_{m,k}[t]$ & User association indicator &
$\mathbf{A}[t]$ & User association matrix \\

$\mathbf{a}_m[t]$ & Action of UAV $m$ &
$A_r$ & Rotor disc area &
$B$ & System bandwidth \\

$c_1, c_2$ & PSO Acceleration coefficients &
$\mathbf{C}_m$ & TAKM cluster centroid $m$ &
$C_m[t]$ & Total power consumption of UAV $m$ \\

$d_{m,k}[t]$ & Distance between UAV $m$ and user $k$ &
$d_{\min}$ & Minimum separation distance &
$d_{\mathrm{safe}}$ & Safety margin around threat zones \\

$d_f$ & Fuselage drag ratio &
$D_n$ & Minimum distance to threat $n$ &
$\mathbf{g}_n$ & Center of threat zone $n$ \\

$g_{m,k}[t]$ & Channel power gain &
$H_m$ & Altitude of UAV $m$ &
$I$ & Max iterations in TAKM \\

$K$ & Number of GUs &
$L_a, L_c$ & Layers in Actor/Critic &
$M$ & Number of UAVs \\

$N$ & Number of threat zones &
$N_a^{(i)}, N_c^{(j)}$ & Neuron counts in layers &
$N_m$ & Number of users served by UAV $m$ \\

$\mathbf{o}_m[t]$ & Observation of UAV $m$ &
$p_m[t]$ & Transmit power of UAV $m$ &
$P(v[t])$ & UAV propulsion power \\

$P_0$ & Blade profile power &
$P_i$ & Induced power in hovering &
$P_{\max}$ & Maximum UAV transmit power \\

$P_{\mathrm{saf}}[t]$ & Safety violation indicator &
$\pi_m(\cdot)$ & Policy of UAV $m$ &
$\mathbf{q}_k$ & Horizontal position of user $k$ \\

$Q_m(\cdot)$ & Action-value function &
$r[t]$ & Global reward &
$r_n$ & Radius of threat zone $n$ \\

$R_k[t]$ & Data rate of user $k$ &
$R_{\min}$ & Minimum required data rate &
$\mathbf{s}[t]$ & Global environment state \\

$s_r$ & Rotor solidity &
$t$ & Time-slot index &
$t_{\max}$ & Total number of time slots \\

$\mathbf{u}_m[t]$ & Horizontal position of UAV $m$ &
$U_{\mathrm{tip}}$ & Rotor tip speed &
$v_0$ & Mean rotor induced velocity \\

$V_{\max}$ & Maximum UAV speed &
$\omega$ & PSO inertia weight &
$\xi_1, \xi_2$ & PSO random variables \\

$\zeta_e, \zeta_c, \zeta_{\mathrm{saf}}$ & Reward weighting factors &
$\alpha$ & Learning rate &
$\gamma$ & Discount factor \\

$\gamma_{\mathrm{LoS}}$ & Additional loss for LoS &
$\gamma_{\mathrm{NLoS}}$ & Additional loss for NLoS &
$\delta_t$ & Duration of a time slot \\

$\eta_{m,k}[t]$ & SINR between UAV $m$ and user $k$ &
$\theta_m^\pi$ & Policy network parameters &
$\rho$ & Air density \\

$\sigma_e^2$ & Exploration noise variance &
$\sigma_n^2$ & Noise power &
$\tau$ & Target network update factor \\
\hline
\end{tabular}
\vspace{-1em}
\end{table*}

\section{System Model}

\begin{figure}[t]
	\centering
	\includegraphics[width=3.5 in]{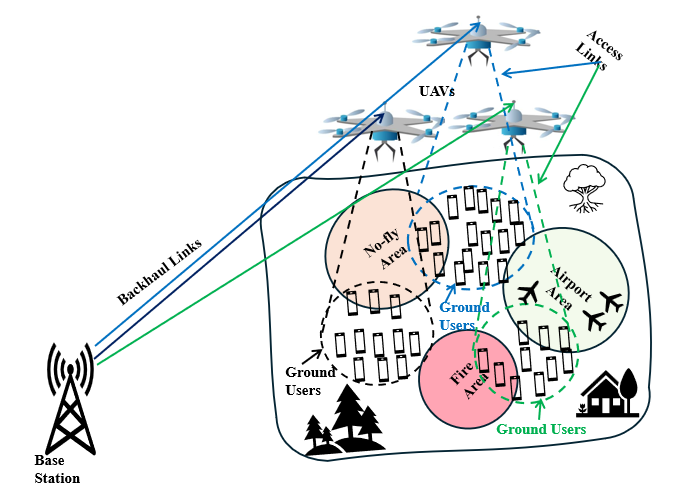}
	\vspace{-1em}
	\caption{The proposed scenario.}
	\label{fig_1}
	\vspace{-1.5em}
\end{figure}

We consider a downlink multi-UAV-assisted wireless communication system consisting of $M$ UAVs, represented by the set $\mathcal{M}=\{1,2,\ldots,M\}$, and $K$ GUs, represented by the set $\mathcal{K}=\{1,2,\ldots,K\}$. The UAVs deliver mission-critical traffic in a disaster area, including voice calls and messaging, with a guaranteed minimum data rate per user. The GUs are non-uniformly distributed to represent a stabilized post-disaster snapshot. As illustrated in the 3D plot in Section VI, this model captures a scenario where a small number of users are trapped within hazardous zones while the majority have concentrated into emergency assembly points. Thus, the stationary assumption is reasonable for both the trapped users and the assembly points.  The UAVs act as aerial base stations to provide wireless service to the GUs over a time horizon 
$\mathcal{T} = \{1, 2, \dots, t_{\max}\}$. The proposed scenario, illustrated in Fig.~\ref{fig_1}, includes a ground base station that establishes a backhaul connection to the UAVs. The horizontal coordinates of UAV 
$m$ at time step $t$ are represented by 
$\mathbf{u}_m[t] = (x_m[t], y_m[t])$, with a fixed altitude 
$H_m$. The horizontal position of GU 
$k$ is denoted by 
$\mathbf{q}_k = (x_k, y_k)$. The GUs are assumed to be stationary, and each user is  served by a single UAV using frequency-division multiple access (FDMA).

In addition to the communication links, the environment contains 
$N$ threat zones, represented by 
$\mathcal{N} = \{1, 2, \dots, N\}$, each characterized by its center coordinate 
$\mathbf{g}_{n}= (x_n, y_n)$ and radius 
$r_{n}$. These zones may represent hazards such as fire areas, restricted or no-fly regions, and other dangerous zones, and UAVs must avoid entering them for safety reason. Although UAVs are strictly prohibited from entering threat zones, GUs  may be located both inside and outside these areas. Consequently, UAV trajectories must be designed to ensure coverage for all users while keeping the UAVs outside the boundaries of the threat zones.

The UAV-GUs association at time step $t$ is represented by a binary matrix 
$\mathbf{A}[t] = [a_{m,k}[t]] \in \{0,1\}^{M \times K}$, where 
$a_{m,k}[t] = 1$ if UAV $m$ serves GU $k$ at time $t$, and 
$a_{m,k}[t] = 0$ otherwise. Each GU is served by exactly one UAV per time slot:
\begin{equation}
    \sum_{m=1}^{M} a_{m,k}[t] = 1, \quad \forall k \in \mathcal{K}, \; t \in \mathcal{T}.
\end{equation}

\noindent The  distance between the  UAV \textit{m} and the GU \textit{k} is given by:
\begin{equation}
d_{m,k}[t] = \sqrt{\|\mathbf{u}_m[t] - \mathbf{q}_k\|^2 + H_m^2},
\label{eq:distance}
\end{equation}
where $\left\| \cdot \right\|$ denotes the Euclidean norm, and $\|\mathbf{u}_m[t] - \mathbf{q}_k\|$ represents the horizontal distance between the UAV \textit{m} and GU \textit{k} at time step \textit{t}. The UAV trajectory must satisfy several safety constraints. First, the maximum speed constraint is imposed as
\begin{equation}
\| \mathbf{u}_m[t+1] - \mathbf{u}_m[t] \| \leq V_{\max}\delta_t,\quad \forall m\in \mathcal{M},\quad  t \in \mathcal{T},
\end{equation}
where $V_{\max}$ denotes the UAVs’ maximum allowable speed, and $\delta_t$ is the duration of the time slot. $\mathbf{u}[t]$ and $\mathbf{u}[t+1]$ are the UAV positions at time steps $t$ and $t+1$, respectively. To prevent collisions, the minimum separation between any two UAVs is enforced by
\begin{equation}
\| \mathbf{u}_i[t] - \mathbf{u}_j[t] \| \geq d_{\min},\quad \ \forall i \neq j,\quad  t \in \mathcal{T},
\end{equation}
with $d_{\min}$ is the minimum distance between the UAVs to avoid collision. In addition, the trajectories must avoid threat zones by maintaining a safe horizontal distance from each threat location, expressed as
\begin{equation}
\| \mathbf{u}_m[t] - \mathbf{g}_n \| \geq D_n, \quad \forall m \in \mathcal{M},\; n \in \mathcal{N},\; t \in \mathcal{T},
\end{equation}
where $D_n$ represents the minimum allowable horizontal distance to threat zone $n$. The distance $D_n$ is given by $D_n = r_n + d_{safe}$, with $d_{safe}$ representing the safety margin.

The UAV-to-ground channel is modeled using a probabilistic pathloss formulation that accounts for both LoS and non-LoS (NLoS) links, each occurring with different probabilities \cite{7037248, 6863654}. The probability of a LoS connection between the UAV \textit{m} and the GU \textit{k} is given by  \cite{6863654}:
\begin{equation}
P_{m,k}^{\text{LoS}}[t] = \frac{1}{1 + a \exp\left( -b \left( \frac{180}{\pi} (\theta_{m,k}[t])-a\right]\right\}},
\end{equation}
where \(a\) and \(b\) are constants that depend on the environment. The elevation angle \(\theta_{m,k}[t]\) from GU \( k \) to UAV \( m \) is given by
\begin{equation}
\theta_{m,k}[t] = \tan^{-1}\left(\frac{H_m}{\|\mathbf{u}_m[t] - \mathbf{q}_k\|}\right).
\end{equation}

 The probability of NLoS propagation is given by \(P_{m,k}^{\text{NLoS}}[t] = 1 -P_{m,k}^{\text{LoS}}[t]\). The mean path loss using the LoS and NLoS between UAV \textit{m} and GU \textit{k} at time step \textit{t} can be expressed as
\begin{equation}
L_{m,k}[t] = L_{m,k}^{\text{LoS}}[t] \times P_{m,k}^{\text{LoS}}[t] + L_{m,k}^{\text{NLoS}}[t] \times P_{m,k}^{\text{NLoS}}[t],
\label{eq:mean_path_loss}
\end{equation}
where $L_{m,k}^{\text{LoS}}$ and $L_{m,k}^{\text{NLoS}}$ denote the pathloss models, in dB, for the LoS and NLoS links, respectively, between UAV~$m$ and GU~$k$, and are given by \cite{7037248,6863654}:

\begin{equation}
L_{m,k}^{\text{LoS}}[t] = 20 \log\left(\frac{4 \pi f_c d_{m,k}[t]}{c}\right) + \gamma_{\text{LoS}},
\label{eq:LoS_path_loss}
\end{equation}
and
\begin{equation}
L_{m,k}^{\text{NLoS}}[t] = 20 \log\left(\frac{4 \pi f_c d_{m,k}[t]}{c}\right) + \gamma_{\text{NLoS}},
\label{eq:NLoS_path_loss}
\end{equation}
where \( f_c \) is the carrier frequency, \( c \) is the speed of light. \( \gamma_{\text{LoS}} \) and \( \gamma_{\text{NLoS}} \) are the additional losses for LoS and NLoS, respectively. 

The channel power gain between UAV $m$ 
and user $k$ at time slot $t$ can be modeled as \cite{8727504} 
\begin{equation}
g_{m,k}[t] = \left( \frac{c}{4 \pi f_c d_{m,k}[t]} \right)^{2}\left(P^{\text{LoS}}_{m,k}[t] \, \gamma_{\text{LoS}} 
+ P^{\text{NLoS}}_{m,k}[t] \, \gamma_{\text{NLoS}}\right)^{-1},
\end{equation}
where the first bracket denotes the free-space pathloss term, accounting for distance-dependent attenuation, and the second bracket represents the shadowing factor under LoS and NLoS conditions.
As each UAV is allocated an orthogonal channel, co-channel interference can be considered negligible. Consequently, when GU $k$ is served by UAV $m$ at time step $t$, the corresponding signal-to-interference-plus-noise ratio (SINR) is expressed as:
\begin{equation}
\eta_{m,k}[t] = \frac{g_{m,k}[t]a_{m,k}[t] p_m[t] }{\sigma_n^2 },
\label{eq:sinr}
\end{equation}
where $p_m[t]$ is the transmit power of UAV~$m$ at time step~$t$, and it satisfies $0 \leq \sum_{k=1}^{K} a_{m,k}[t]\,p_m[t] \leq P_{\max},$ where $P_{\max}$ denotes the maximum allowable transmit power of a UAV. The term $\sigma_n^2$ denotes the noise power. 
In this paper, a single power level is assumed for each UAV~$m$, which is equally shared among all its associated users.  $a_{m,k}[t]$ denotes the binary user association indicator.

Since UAVs are allocated orthogonal channels, the inter-UAV co-channel interference can be neglected.
Accordingly, the achievable data rate of GU~$k$ at time step~$t$ is given by
\begin{equation}
R_k[t] = \sum_{m=1}^M (B/N_m) \log_2 \left( 1 + \eta_{m,k}[t] \right),
\end{equation}
where \( B \) is the total bandwidth and \( N_m \) is the number of users served by UAV \( m \).

We consider a power consumption model that includes both the communication power for transmission and the propulsion power for UAV mobility, as in \cite{9826431}. Accordingly, the total power consumption of UAV $m$ at time step $t$ is expressed as:
\begin{equation}
C_m[t] =  \sum_{k=1}^K a_{m,k}[t]\, p_m[t] + P(v[t]) ,
\end{equation}
where $P(v[t])$ denotes the flight power of the UAV as a function of its speed. The flight power consumption model $P(v[t])$ for rotary-wing UAV is given by \cite{8663615}:
\begin{equation}
\begin{split}
P(v[t]) &= P_0 \left(1 + \frac{3 v^2[t]}{U_{\text{tip}}^2}\right) 
+ P_i \sqrt{1 + \frac{v^4[t]}{4 v_0^4} - \frac{v^2[t]}{2 v_0^2}} \\
&\quad + \frac{1}{2} d_f \rho s_r A_r v^3[t],
\label{eq:flight_power}
\end{split}
\end{equation}
where $P_0$ denotes the blade profile power, and $P_i$ corresponds to the induced power during hovering. The tip speed of the rotor blade is represented by $U_{\mathrm{tip}}$, while $v_0$ indicates the mean velocity induced by the rotor. The parameters $d_f$, $\rho$, $s_r$, and $A_r$ denote the fuselage drag ratio, air density, rotor solidity, and rotor disc area, respectively.

\section{Problem Formulation}
Our aim is to provide the GUs with high QoS while minimizing the energy consumption of the UAVs. To this end, we formulate a joint optimization problem that simultaneously considers the number of UAVs, user association, trajectory planning, and transmit power allocation. The primary objective is to maximize the global energy efficiency (EE) of the system, defined as the sum rate of all GUs over the total energy consumed by all UAVs. 

The proposed optimization is subject to critical operational and safety constraints, including guaranteeing full coverage and a minimum QoS rate for all users, strictly avoiding UAV collisions, prohibiting entry into designated threat zones with a specified safety margin, and satisfying maximum UAV velocity and transmit power limits. The proposed problem can be formulated as: 
\begin{subequations}\label{eq:optimization}
\begin{align}
\max_{M,\,\mathbf{A[t]},\,\mathbf{Q},\,\mathbf{P}} \quad 
& \frac{1}{t_{\max}} \sum_{t=1}^{t_{\max}} \Bigg[
  \frac{\sum_{k=1}^K R_k[t]}{\sum_{m=1}^M C_m[t]} \Bigg] \tag{16}\label{eq:objective_main} \\
\text{s.t.} \quad \quad
& a_{m,k}[t] \in \{0,1\}, \quad \quad \quad \quad \quad \ \forall m, k , t , \label{eq:cons_1} \\
& \sum_{m=1}^{M} a_{m,k}[t] = 1, \quad \quad \quad \quad \quad \quad \ \forall k , t, \label{eq:cons_2} \\
& \|\mathbf{u}_m[t+1] - \mathbf{u}_m[t]\| \leq V_{\max} \delta_t, \  \forall m, t, \label{eq:cons_3} \\
& \|\mathbf{u}_i[t] - \mathbf{u}_j[t]\| \geq d_{\min}, \ \  \forall i,j,t, i \neq j,\;  \label{eq:cons_4} \\
& 0 \leq \sum_{k=1}^{K} a_{m,k}[t] p_m[t] \leq P_{\max},  \quad \forall m , t, \label{eq:cons_5} \\
& \|\mathbf{u}_m[t] - \mathbf{g}_{n}\| \geq D_{n}, \quad \quad \quad \  \forall m, n, t, \label{eq:cons_6} \\
& x_{\min} \leq x_m[t] \leq x_{\max}, \quad  \quad \quad \quad \forall m, t, \label{eq:cons_7} \\
& y_{\min} \leq y_m[t] \leq y_{\max}, \quad  \quad \quad \quad \ \forall m, t, \label{eq:cons_8} \\
& R_k[t] \geq R_{\min}, \quad  \quad \quad  \quad \quad  \quad \quad \quad \forall k, t, \label{eq:cons_9}
\end{align}
\end{subequations}
where the matrix $\mathbf{Q}$, of size $M \times t_{\text{max}}$, represents the trajectories of all $M$ UAVs, with each element corresponding to the 2D horizontal coordinates of UAV $m$ at time step $t$. The matrix $\mathbf{P}$, of dimension $M \times t_{\text{max}}$, denotes the transmit power allocation for all UAVs, with each entry $p_m[t]$ specifying the transmission power level of UAV $m$ at time step $t$. It is important to note that the EE metric considers only UAV power, since the grid-powered base station is not energy-limited like the UAVs.

The objective function in \textcolor{blue}{\eqref{eq:optimization}} is subject to a set of operationally realistic constraints. User association is enforced through binary assignment variables \eqref{eq:cons_1} and exclusive connectivity requirements \eqref{eq:cons_2}. UAV mobility is restricted by maximum velocity limits \eqref{eq:cons_3} and collision-avoidance constraints \eqref{eq:cons_4}. Power allocation adheres to per-UAV transmit limits \eqref{eq:cons_5}. Operational safety is maintained by enforcing threat-zone avoidance \eqref{eq:cons_6} and adherence to geographical boundaries \eqref{eq:cons_7}–\eqref{eq:cons_8}. Finally, constraint \eqref{eq:cons_9} ensures that all users receive a guaranteed minimum quality of service. Collectively, these constraints establish a formulation that integrates system performance with the practical requirements of real-world deployment.

The optimization problem  in \textcolor{blue}{\eqref{eq:optimization}} is a mixed-integer non-linear  programming problem (MINLP), characterized by its combinatorial nature due to the binary user association variables in $\mathbf{A}$ (constraint \eqref{eq:cons_1}) and the continuous variables in both the trajectory set $\mathbf{Q}$ and power allocation matrix $\mathbf{P}$. The objective function, which represents the EE of the system, is a fractional and non-convex function, further complicating the optimization process. Additionally, constraints such as the non-convex collision avoidance requirement in \eqref{eq:cons_4}, the threat-zone avoidance in \eqref{eq:cons_6}, and the minimum rate requirement in \eqref{eq:cons_9} contribute to the overall non-convexity of the problem. These attributes collectively make \textcolor{blue}{\eqref{eq:optimization}} NP-hard and computationally challenging to solve using conventional optimization techniques.

To address these challenges, the problem in \textcolor{blue}{\eqref{eq:objective_main}} is decomposed into three subproblems. First, the required number of UAVs is determined and the proposed TAKM algorithm is applied to establish the initial user associations, and generate an initial safe UAV placement that satisfies geographical constraints. Second, UAVs are deployed to the identified initial clusters by minimizing the total flight distance and positioned at the cluster centroids. This pre-placement strategy narrows the search space and accelerates the convergence of the learning process. Finally, a threat-aware MARL scheme is proposed to optimize user associations, UAVs trajectories and power control.

\section{Clustering and UAV Matching}

\subsection{Threat-Avoidance K-means Clustering Algorithm}

The conventional $K$-means clustering algorithm has been widely adopted for UAV placement due to its computational efficiency and effectiveness in minimizing the total within-cluster variance. The algorithm iteratively performs two main steps: (1) assignment of users to their nearest centroid (UAV), and (2) recalculation of centroids as the mean position of assigned users. This process converges to a locally optimal solution that minimizes the sum of squared Euclidean distances between users and their serving UAVs. However, CKM algorithm suffers from a significant drawback in threat-prone environments: the computed mean positions may fall within prohibited threat zones, violating safety constraints. The CKM algorithm lacks any mechanism to incorporate geographical constraints, making it unsuitable for real-world operations where airspace restrictions must be strictly observed. To address these limitations, we introduce a  TAKM algorithm, which integrates threat avoidance mechanisms into the CKM algorithm through the following key enhancements:
\begin{enumerate}
    \item {Safe Initialization}: The TAKM algorithm begins by defining threat buffers around each threat zone, extending the original radius by a safety margin $d_{safe}$. A comprehensive grid scan $\mathcal{S}$ is performed across the operational area to precompute all safe positions that lie outside all threat buffers. Initial centroids are then randomly selected exclusively from this set of safe positions, ensuring the optimization process starts from feasible solutions. 
    \item {Constrained Centroid Update}: During each iteration, after computing the conventional mean position $\mathbf{\mu}_m$ for each cluster, the algorithm performs a safety verification check:
 \begin{equation}
\text{IsSafe}(\mu_m) = 
\begin{cases} 
\text{true}, & \text{if } \|\mathbf{\mu}_m - \mathbf{g}_m\| > r_n + d_{safe} \\[-2pt]
             & \quad \forall n = 1,\dots,N, \\[6pt]
\text{false}, & \text{otherwise}.
\end{cases}
\end{equation}

If the mean position is safe, it is accepted as the new centroid. If unsafe, the algorithm projects the mean to the nearest safe position:

\begin{equation}
\mathbf{C}_m^{(i+1)} = \arg\min_{\textit{p} \in \mathcal{S}} \|p - \mathbf{\mu}_m\|
\end{equation}

This projection ensures feasibility while minimizing deviation from the optimal cluster mean.

\end{enumerate}

The proposed TAKM algorithm maintains the convergence properties of the CKM algorithm while ensuring all intermediate and final solutions satisfy the safety constraints. The termination condition considers both the stabilization of centroid positions and the maintenance of safety guarantees. The complete pseudocode of the proposed TAKM clustering is shown in Algorithm~\ref{alg:threat_aware_kmeans}.

\begin{algorithm}[t]
\caption{The TAKM Clustering Algorithm}
\label{alg:threat_aware_kmeans}
\begin{algorithmic}[1]
\STATE \textbf{Input:} $M$ (number of clusters), $K$ (number of GUs), GUs positions, threat zones $\{(\mathbf{g}_{n}, r_n)\}_{n=1}^N$, safety margin $d_{safe}$, area boundaries, $\epsilon$
\STATE \textbf{Step 1: Safe Initialization}

\STATE Create a 2D grid of points over the area
\FOR{each threat $n = 1,\dots,N$}
    \STATE Mark grid points inside $r_n + d_{safe}$ of threat $n$ as unsafe
\ENDFOR
\STATE Randomly select $M$ safe positions as initial cluster centroids  $\{\mathbf{C}_m\}_{m=1}^M$
\STATE \textbf{Step 2: Constrained Clustering}
\FOR{iteration $i = 1$ to $I_\text{max}$}
    \STATE Assign each GU to the nearest cluster centroid
    \STATE Store previous centroids for convergence check: 
     $\{\mathbf{C}_m^{prev}=\mathbf{C}_m\}_{m=1}^M$
    \FOR{each cluster $m = 1,\dots,M$}
        \STATE Compute the mean of assigned GUs: $\mathbf{\mu}_m$
        \IF{$\mathbf{\mu}_m$ is outside all threat zones + safety margin}
            \STATE Update centroid $\mathbf{C}_m \gets \mathbf{\mu}_m$
        \ELSE
            \STATE Project $\mathbf{\mu}_m$ to nearest safe position from precomputed safe positions
        \ENDIF
    \ENDFOR
    \STATE \textbf{Check convergence}
    \FOR{each cluster $m = 1,\dots,M$}
    \STATE Compute: $\Delta_m = \|\mathbf{C}_m - \mathbf{C}_m^{prev}\|$
    \ENDFOR
    \IF{$\{\Delta_m\}_{m=1}^M < \epsilon$} \STATE break \ENDIF
\ENDFOR
\STATE \textbf{Return:} Centroid positions $\{\mathbf{C}_m\}_{m=1}^M$  
\end{algorithmic}
\end{algorithm}

\subsection{Joint Determination of the Minimum Number of UAVs and Initial User Association}

Determining the minimum number of UAVs is a critical step for cost-effective deployment,  but existing approaches often assume a fixed number. To address this limitation, we introduce a systematic, constraint-aware procedure that guarantees feasibility while minimizing deployment costs. As outlined in Algorithm~\ref{alg:adaptive_takm}, our method iteratively applies the TAKM clustering algorithm (Algorithm~\ref{alg:threat_aware_kmeans}), starting with \(M = 1\) UAV and incrementing \(M\) until the resulting cluster configuration satisfies all coverage, minimum rate, and threat-avoidance constraints. This process simultaneously yields the minimum number of  UAVs  \(M\), the initial safe UAV positions, and the corresponding initial user association matrix \(\mathbf{A}\).

A key constraint ensures that each GU achieves at least the minimum data rate $R_{\text{min}}$. For each candidate clustering, the per‑user transmit power is set to $p_{m} = P_{\max}/N_{m}$. The achievable rate $R_{k}$ is then computed using the SINR expression in (12) and the rate formula (13). A clustering is valid only if $R_{k} \geq R_{\text{min}}$ holds for every user $k$.

Algorithm~\ref{alg:adaptive_takm} repeats this verification while incrementing $M$ until the rate requirement  is met for all users while ensuring UAV safety. The output of this algorithm is the minimum number of UAVs $M$, the initial safe UAV positions $\{\mathbf{C}_m\}_{m=1}^{M}$ (obtained via TAKM), and the corresponding user association matrix $\mathbf{A}$.
The complete procedure is detailed in Algorithm~\ref{alg:adaptive_takm}.

\subsection{UAV Matching}
Following the execution of Algorithm~\ref{alg:adaptive_takm}, which outputs the safe target cluster centroids, each physical UAV must be assigned to a specific location. Since the UAVs are homogeneous but may be initially dispersed, this task is treated as a minimum total distance assignment problem. The primary objective is to minimize the cumulative Euclidean distance from the UAVs' initial launch positions to their designated centroids, thereby ensuring the most energy-efficient initial deployment possible.

To achieve this, we adopt an optimal matching scheme from the literature based on the Hungarian algorithm (also known as the Kuhn-Munkres algorithm). This method solves the assignment task in polynomial time, guaranteeing a one-to-one mapping that minimizes the global movement cost, thereby making the system more energy-efficient during the deployment phase.

\section{Threat-Aware MARL for user association, trajectory design and power control}

Once the minimum required number of UAVs ($M$) and their initial safe placements are determined in Section IV, the original optimization problem in \eqref{eq:objective_main} can be refined for a machine training/learning phase. In this stage, $M$ is no longer a decision variable but a fixed parameter. The objective remains to maximize the system’s global EE by optimizing the remaining variables: the user association matrix $\mathbf{A}[t]$, the trajectory set $\mathbf{Q}$, and the power allocation matrix $\mathbf{P}$. 

The resulting problem is nonlinear and nonconvex, making traditional optimization methods intractable. Therefore, a MARL framework is proposed in this section to obtain a feasible solution. Moreover, the problem constitutes a hybrid optimization problem involving discrete user association variables $\mathbf{A}[t]$ and continuous power and trajectory variables, $\mathbf{P}[t]$ and $\mathbf{Q}[t]$. Due to this mixed structure, the problem cannot be treated purely as a continuous control problem or solely as a discrete assignment problem; instead, it requires a solution capable of handling both types of variables simultaneously.

\subsection{Markov Decision Process Formulation}

The movement of UAVs directly impacts the data rate of GUs as well as the allocated transmission power. Hence, the actions chosen by UAVs affect both the immediate outcome and the future state of the system. This interaction can be effectively modeled through a partially observable Markov decision process (POMDP), providing a rigorous framework for sequential decision-making in environments characterized by uncertainty.

The reinforcement learning parameter settings can be detailed as follows:

\begin{algorithm}[t]
\caption{Joint Determination of the Minimum Number of UAVs and Initial User Association}
\label{alg:adaptive_takm}
\begin{algorithmic}[1]
\STATE \textbf{Input:} $K$, GU positions, threat zones $\{(\mathbf{c}_{n}, r_n)\}_{n=1}^N$, $d_{safe}$, area boundaries, $P_{\max}$, $\sigma_n^2$, $B$, $R_{\min}$, channel parameters.
\STATE $M \gets 1$

\REPEAT
    \STATE Apply TAKM (Algorithm~\ref{alg:threat_aware_kmeans}) to obtain centroids $\{\mathbf{C}_m\}$ and $\mathbf{A}$
    
    \STATE all\_covered $\gets$ true
    \FOR{each user $k = 1,\dots,K$}
        \STATE Find the number of served users $N_m$ for the UAV $m$ serving user $k$
        \STATE Compute the transmit power per user: $p_{m} = P_{\max}/N_m$
        \STATE Compute channel gain $g_{m,k}$ using (12) %
        \STATE Compute SINR $\eta_{m,k}$ using (13)
        \STATE Compute achievable rate $R_{k}$ using (14)
        
        \IF{$R_{k} < R_{\min}$}
            \STATE all\_covered $\gets$ false
            \STATE \textbf{break}
        \ENDIF
    \ENDFOR

    \IF{not all\_covered}
        \STATE $M \gets M + 1$
    \ENDIF
\UNTIL{all\_covered}
\STATE \textbf{Return:} $M$, $\mathbf{A}$, $\{\mathbf{C}_m\}_{m=1}^M$ 
\end{algorithmic}
\end{algorithm}

\begin{figure*}[!t]
\centering
\includegraphics[width=7 in]{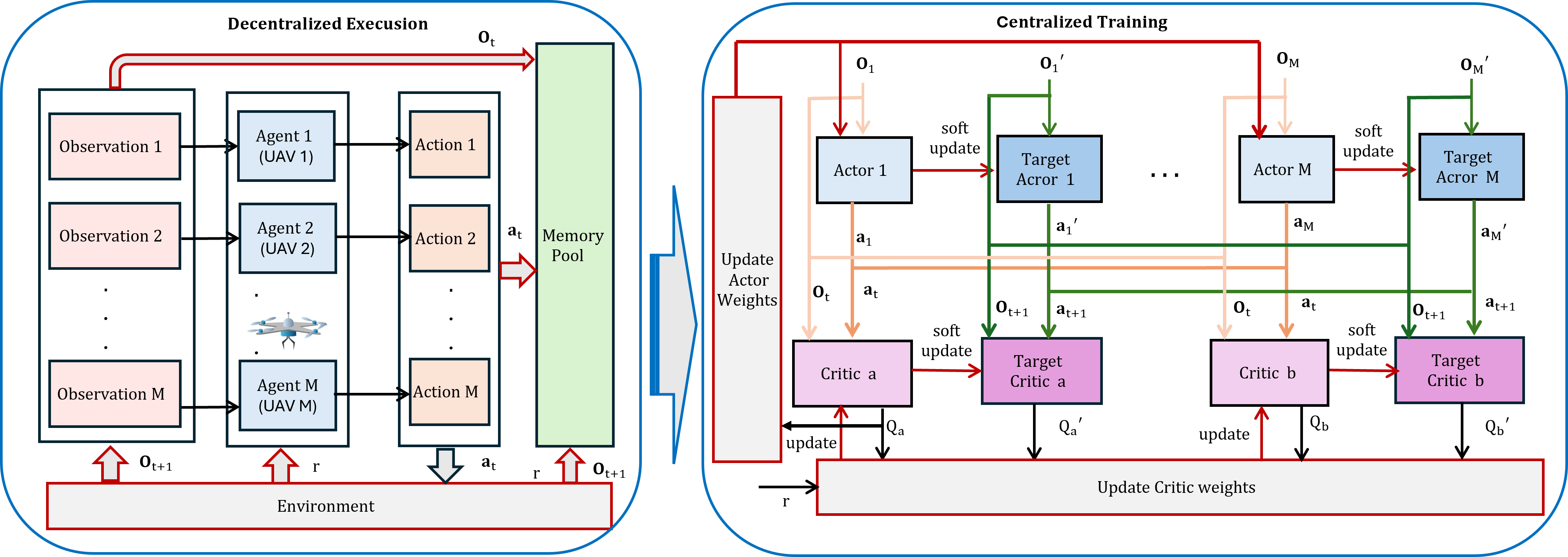}
\caption{ Threat-aware MATD3 structure.}
\label{MATD3_struct}
\vspace{-1em}
\end{figure*}

\begin{itemize}
    \item \textit{State:} The state comprises the coordinates of all UAVs. It is a continuous variable, with bounds determined by the dimensions of the service area. 
\item \textit{Observation:} Each agent can only observe a subset of the state. For UAV $m$, only its own horizontal coordinate is observable, i.e., $\textbf{o}_m = \{ \textbf{q}_m \}$. As a subset of the state, the observation is also a continuous and finite variable, with bounds defined by the coordinates of the service area.

\item \textit{Action:} The action vector $\mathbf{a}_m[t]$ for UAV $m$ at time step $t$ 
is defined by its flight speed, flight angle, and 
transmit power, i.e.,
\begin{equation}
\mathbf{a}_m[t] = \big[ a_{m}^{(v)}[t], a_{m}^{(\phi)}[t], a_{m}^{(p)}[t] \big{]}.
\end{equation}
Given the flight speed $v_m[t] = a_{m}^{(v)}[t]\cdot V_{\max}$ and 
the flight angle $\phi_m[t] = a_{m}^{(\phi)}[t]\cdot 2\pi$, 
the UAV position is updated as
\begin{align}
x_m[t+1] &= x_m[t] + v_m[t]\cos(\phi_m[t])\cdot \delta_t, \\
y_m[t+1] &= y_m[t] + v_m[t]\sin(\phi_m[t])\cdot \delta_t.
\end{align}
\item \textit{Reward:} We propose the following reward function:
\begin{equation} 
\begin{aligned}
r[t] = & \; \frac{\zeta_e}{E_{\max}} 
\cdot \frac{\sum_{m=1}^{M}\sum_{k=1}^{K} a_{m,k}[t]\, R_{k}[t]}
{\sum_{m=1}^{M} C_m[t]} ~\\
&- \zeta_c \cdot \left[ 1 -\frac{ \sum_{m=1}^{M}\sum_{k=1}^{K} a_{m,k}[t]}{K}\right] - \zeta_{\text{saf}} \cdot P_{\text{saf}}[t],
\end{aligned}
\end{equation}
where $E_{\max}$ is a fixed normalization constant used to scale the EE term in the reward. The variable \(P_{\text{saf}}(t)\) is a binary safety flag that takes the value 1 when a collision occurs, a UAV enters a threat zone, or a UAV flies outside the defined operational boundaries at time step \(t\), and 0 otherwise. The coefficients $\zeta_e$, $\zeta_c$, and $\zeta_{\text{saf}}$ are positive weighting parameters that balance \textit{EE}, \textit{coverage}, and \textit{safety} in the reward function. They were tuned empirically to ensure stable learning and constraint satisfaction, with $\zeta_{\text{saf}}$ set higher to strongly penalize safety violations.

\end{itemize}

The POMDP is formulated to select a policy 
$\pi(\textbf{o}_1, \ldots, \textbf{o}_M)$ that maximizes the cumulative reward, expressed as
\begin{equation}
    G[t] = \sum_{t'=t}^{\infty} \alpha^{\,t'-t} r[t],
    \label{eq:G}
\end{equation}
where $\alpha \in [0,1[$ represents the discount factor that balances immediate and future rewards.  
Given the current observation set $\textbf{o}[t]=\{\textbf{o}_1, \ldots , \textbf{o}_M\}$ and the corresponding action 
$\textbf{a}[t]=\{\textbf{a}_1, \ldots, \textbf{a}_M\}$, the expected cumulative return under the policy 
$\pi(\textbf{o}_1, \ldots, \textbf{o}_M)$ can be formulated as
\begin{equation}
    Q(\textbf{o}[t], \textbf{a}[t]) 
    = \mathbb{E}\left[G\mid\textbf{o}[t], \textbf{a}[t] \right].
    \label{eq:Q}
\end{equation}

 The Bellman equation is then used to determine the optimal action-value function called the Q-value and defined as:
\begin{align}
    Q^{*}(\textbf{o}[t], \textbf{a}[t]) 
    &= r + \alpha \max_{\textbf{a}[t+1]} 
    Q^{*}(\textbf{o}[t+1], \textbf{a}[t+1]),
    \label{eq:Bellman}
\end{align}
where  $\textbf{o}[t+1]=\{\textbf{o}_1', \ldots, \textbf{o}_M'\}$ and  $\textbf{a}[t+1]=\{\textbf{a}_1', \ldots, \textbf{a}_M'\}$ correspond to the subsequent observations and actions.  
An important point is that the maximization in \eqref{eq:Bellman} can introduce a bias, often resulting in an overestimated Q-value.

\subsection{Multi-Agent RL Method}

Since the observations and actions in our UAV placement and resource allocation problem are continuous variables, conventional tabular methods (e.g., Q-learning) are not suitable for solving the optimization task. To address this, MARL algorithms such as MADDPG and MATD3 have been proposed for continuous state-action spaces. Among these, we employ MATD3 because it effectively mitigates the Q-value overestimation problem inherent in MADDPG through the use of double critics and delayed policy updates.

To maintain stability and avoid the complexity of hybrid action spaces \cite{11391547}, we decompose the problem into a continuous control task and a deterministic assignment sub-problem. Because MATD3 handles only continuous actions, the agents do not explicitly optimize the association variables through direct action outputs. Instead, the associations are determined deterministically by the environment based on the UAV positions and transmit power levels. Since the global reward is computed after the association update, the agents indirectly learn deployment and power-control strategies that produce favorable user associations. Therefore, the framework learns the effects of the association mechanism implicitly through reward feedback, rather than explicitly learning the association matrix itself. 

The overall architecture of the proposed threat-aware MATD3 is illustrated in Fig.~\ref{MATD3_struct}.
\begin{algorithm}[t]
\caption{The Proposed MATD3 with TAKM clustering and pre-placement}
\label{alg:MATD3}
\begin{algorithmic}[1]
\STATE \textbf{Input:} Actor and critic network structures; UAV initial placement; exploration variance $\sigma_e^2$; decay factor $\delta$; environment parameters; GUs positions; threat zones $\{(\mathbf{g}_{n}, r_n)\}_{n=1}^N$; parameters in  Table~\ref{table:SimulationParameters}
\STATE Initialize actor network weights $\mathbf{U}_m$ and critic network weights $\mathbf{W}_j, \; j\in\{1,2\}$
\STATE Initialize target networks: $\mathbf{U}'_m \leftarrow \mathbf{U}_m$, $\mathbf{W}'_j \leftarrow \mathbf{W}_j$
\STATE Initialize replay memory $\mathcal{D} \leftarrow \emptyset$

\FOR{episode $=1,2,\ldots,L$}
    \STATE Initialize UAV positions and exploration noise
    \FOR{$t = 1,2,\ldots,t_{\max}$}
        \STATE \textbf{Each UAV agent} observes its local state $\textbf{o}_m$
        \STATE Select exploratory action with noise:
        \[
            \textbf{a}_m[t] = \mathrm{clip}\big(\pi_{\mathbf{U}_m}(o_m) + \mathcal{N}(0,\sigma_e^2), 0, 1\big)
        \]
        \STATE Execute joint action $\textbf{a}[t]$, observe next state $\textbf{o}'[t]$
        \STATE \textbf{User Association:} Update $\mathbf{A}[t]$ based on current UAV positions and channel conditions
        \STATE Check constraints (collision, threat-zone, operational boundaries, rate requirements)
        \STATE Compute immediate \textbf{global reward} $r$ using~(22) \COMMENT{Shared among all UAV agents}
        \STATE Store transition $(\mathbf{o}[t], \mathbf{a}[t], r, \mathbf{o}[t+1])$ in replay memory $\mathcal{D}$
        
        \IF{$\mathcal{D}$ has sufficient samples }
            \STATE Sample mini-batch of experiences from $\mathcal{D}$
            \STATE \textbf{Critic Update:} Update $\mathbf{W}_1,\mathbf{W}_2$ using gradient descent on~(29)
            \STATE \textbf{Actor Update (delayed):} If policy update condition met, update $\mathbf{U}_m$ using~(30)
            \STATE \textbf{Target Update:} Soft-update target networks $\mathbf{U}'_m$, $\mathbf{W}'_1$, $\mathbf{W}'_2$ using~(32)
        \ENDIF
    \ENDFOR
    
    \STATE Anneal exploration variance: $\sigma_e^2 \leftarrow \delta \sigma_e^2$
\ENDFOR

\STATE \textbf{Return:} $\mathbf{U}_m, \mathbf{W}_1, \mathbf{W}_2$ 
\end{algorithmic}
\end{algorithm}

The learning process in MATD3 follows the centralized training and decentralized execution (CTDE) paradigm. Each UAV agent maintains its own deterministic policy $\pi_{\mathbf{U}_m}(\mathbf{o}_m)$, parameterized by the actor weights $\mathbf{U}_m$, which maps its observation $\mathbf{o}_m$ to a continuous action vector $\mathbf{a}_m$. During training, a centralized critic evaluates the joint state–action pair $(\mathbf{o}[t], \mathbf{a}[t])$. The critic networks are parameterized by the weight vectors $\mathbf{W}_j$ $(j=1,2)$, corresponding to the two critics in the twin delayed architecture. The vectors $\mathbf{U}'_m$ and $\mathbf{W}'_j$ $(j=1,2)$ denote the target actor and target critic weights, respectively.

During experience collection, agents balance exploitation and exploration by following the current policy with some probability and adding noise to encourage exploration. This is given~by
\begin{equation}
\textbf{a}_m = \text{clip} \Big( \pi_{\mathbf{U}_m}(\textbf{o}_m) + \mathcal{N}(0, \sigma_e^2), \; 0, \; 1 \Big),
\end{equation}
where $\sigma_e^2$ is the variance of the exploration noise, $\pi_{\mathbf{U}_m}$ is the policy, and the clip function bounds actions within $[0,1]$.

While the observation space $o_m[t]$ is restricted to local coordinates to ensure communication efficiency during execution, environmental awareness is achieved through a two-tiered safety mechanism. First, the TAKM algorithm provides a safety-aware initial deployment, ensuring agents begin in feasible regions. Second, the CTDE paradigm allows the centralized critics to utilize global state information, including threat boundaries and neighbor positions, to shape the reward signal. By incorporating a binary safety penalty $P_{\text{saf}}$ for constraint violations, the framework implicitly encodes the global environmental geometry into the decentralized policy weights. This allows each agent to navigate complex manifolds and avoid localized threats using only its own state, significantly reducing real-time coordination overhead. This observation design is consistent with \cite{9826431} and is justified for the fixed scenario considered in this work, where users and threat zones are stationary and known.

\subsubsection{Decentralized Execution}
In the execution step, each UAV operates independently using only its local observation. Given an observation $\mathbf{o}_m$, the actor network generates a continuous action vector $\mathbf{a}_m$, corresponding to the UAV’s velocity, flight angle, and transmit power. After executing the action, the UAV interacts with the environment, which produces a new observation and an immediate reward $r$. At every time step the environment deterministically recomputes the association matrix based on the updated UAV positions, transmit power levels, and the channel conditions. This ensures that each user is associated with the most appropriate UAV according to the current system state and guarantees consistency with the rate and coverage constraints. The interaction of all UAVs at time step $t$ produces the joint experience tuple
$(\mathbf{o}[t],\; \mathbf{a}[t],\; r,\; \mathbf{o}[t+1],\; \mathbf{a}[t+1]),
$ which is stored in a replay memory pool. This storage mechanism allows the learning process to reuse past transitions, break temporal correlations, and improve sample efficiency.

\subsubsection{Centralized Training}
During training, a centralized critic is employed to evaluate the joint state-action pair $(\textbf{o}[t], \textbf{a}[t])$, allowing the framework to capture inter-agent dependencies such as collision avoidance. To reduce overestimation bias, MATD3 introduces two critics and computes the target Q-value as
\begin{equation}
Q_t = r + \alpha \min \big\{ Q'_1(\textbf{o}[t+1], \textbf{a}[t+1]), \; Q'_2(\textbf{o}[t+1], \textbf{a}[t+1]) \big\}.
\end{equation}

Training of the critic network is carried out by reducing the squared difference between the $Q_t$ and its evaluated, which is formulated using the mean squared error (MSE) as follows:
\begin{equation}
L(\textbf{W}_j) = \mathbb{E} \Big[ \big(Q_{j}(\textbf{o}[t], \textbf{a}[t]) - Q_t\big)^2 \Big], j=1, 2.
\end{equation}

Then, the critic network weights $\textbf{W}_j$ are updated as
\begin{equation}
\mathbf{W}_j \leftarrow \mathbf{W}_j - \eta \nabla_{\mathbf{W}_j} L(\mathbf{W}_j), 
\quad j = 1,2,
\end{equation}
where $\eta$ is the learning rate. Then, the actor network weights $\textbf{U}_m$ are updated as
\begin{equation}
\mathbf{U}_m \leftarrow \mathbf{U}_m + \eta \nabla_{\mathbf{U}_m} J(\mathbf{U}_m), 
\quad m = 1,...,M,
\end{equation}
where $\nabla_{\mathbf{U}_m} J(\mathbf{U}_m)$ is given by \cite{9826431}:
\begin{equation}
\nabla_{\mathbf{U}_m} J(\mathbf{U}_m) \approx \mathbb{E}\Big[ \nabla_{\mathbf{U}_m}   Q_{1}(\textbf{o}[t], \textbf{a}[t]) \Big].
\end{equation}

\noindent Finally, soft updates are applied to the target networks for both actors and critics  \cite{fujimoto2018td3}:
\begin{equation}
\begin{aligned}
\mathbf{U}'_m &\leftarrow \tau \mathbf{U}_m + (1-\tau)\mathbf{U}'_m,\quad m=1,2, ..., M,\\
\mathbf{W}'_j &\leftarrow \tau \mathbf{W}_j + (1-\tau)\mathbf{W}'_j, \quad j=1,2,
\end{aligned}
\end{equation}
where $\tau \ll 1$ is the soft update factor.  

The combination of delayed policy updates, double critics, and target policy smoothing stabilizes the training process, enabling UAV agents to learn cooperative strategies that maximize system performance while discouraging violations of the safety constraints through the reward-based penalty.  

 In this paper, we employ uniform mini-batch sampling for the experience replay buffer, consistent with \cite{9826431}. The TAKM-based initialization ensures the buffer is quickly populated with high-quality transitions, reducing the 
need for sophisticated prioritization. Nevertheless, advanced techniques such as prioritized experience replay (PER) \cite{schaul2016prioritized} remain 
a promising direction for more complex environments.

The presented MATD3 architecture builds upon the foundational MADDPG framework but introduces critical enhancements to improve learning stability. A key distinction lies in the treatment of the critic networks. In MADDPG, each agent is equipped with its own centralized critic, which is used to compute target Q-values. However, this single-critic design is susceptible to overestimation bias. In contrast, MATD3 employs two centralized critics shared among all agents, and the minimum of their outputs is used to form the target Q-value. This twin-critic mechanism effectively mitigates overestimation by taking their minimum, as defined in Eq. (27). The target Q-value in the MADDPG can be written as:

\begin{equation}
Q_t = r + \alpha Q'\left(\mathbf{o}[t+1], \mathbf{a}[t+1]\right).
\end{equation}

In addition, MATD3 algorithm employs delayed policy updates, where the actor network is updated less frequently than the critics. This mechanism stabilizes the learning process and ensures that target values remain reliable for multi-UAV optimization. The pseudocode of the proposed MATD3 with TAKM and pre-placement is shown in Algorithm~\ref{alg:MATD3}.
The RL agent is trained offline using simulated environments. During deployment, the learned policy is executed online with low computational complexity, as explained in the next section. 

\begin{table*}[htbp]
\centering
\caption{Computational Complexity Comparison of Different Algorithms}
\label{tab:performance_complexity}
\renewcommand{\arraystretch}{1.25}
\begin{tabular}{|p{1.4cm}|p{11cm}|p{4cm}|}
\hline
\textbf{Algorithm} & \textbf{Computational Complexity} & \textbf{Approximate Operations} \\
\hline
MATD3  &  
$\mathcal{O}(MKI) + \mathcal{O}(M^3) + 
\mathcal{O}\!\left(Lt_{\max}\!\left(2M\sum\limits_{i=0}^{L_a} N_a^{(i)}N_a^{(i+1)} + 
4\sum\limits_{j=0}^{L_c} N_c^{(j)}N_c^{(j+1)}\right)\right) + 
\mathcal{O}(Lt_{\max} MK)$ & 
Offline Training: $6.4\times 10^{10}$,  \qquad    Online Deployment: $4.7\times 10^5$ \\ \hline

MADDPG & 
$\mathcal{O}(MKI) + \mathcal{O}(M^3) + 
\mathcal{O}\!\left(Lt_{\max}\!\left(2M\sum\limits_{i=0}^{L_a} N_a^{(i)}N_a^{(i+1)} + 
2\sum\limits_{j=0}^{L_c} N_c^{(j)}N_c^{(j+1)}\right)\right) + 
\mathcal{O}(Lt_{\max} MK)$ & 
Offline Training: $5.8 \times 10^{10}$, \qquad  Online Deployment: $4.7 \times 10^5$ \\ \hline

PSO  & 
$\mathcal{O}(MKI) + \mathcal{O}(M^3) + 
\mathcal{O}\!\left(I_{\text{PSO}} N_p (MK + M^2 + MN)t_{\max}\right)$ & 
Online Deployment: $1.6 \times 10^{10}$ \\ \hline
\end{tabular}
\vspace{-1em}
\end{table*}

\subsection{Meta-heuristic Solution}
To establish a performance benchmark, we implement PSO, a population-based meta-heuristic that solves the joint trajectory and power allocation problem by simulating collective social behavior. It is used solely as a baseline for comparison and is not part of the proposed framework. Each particle $j$ updates its velocity $\mathbf{v}_{j}$ and state $\mathbf{s}_{j}$ based on its personal best, $\mathbf{p}_{{\rm best},j}$, and the swarm’s global best, $\mathbf{g}_{{\rm best}}$, using:
\begin{equation}
\mathbf{v}_{j}[k+1] = \omega \mathbf{v}_{j}[k] + c_1 \xi_1 (\mathbf{p}_{{\rm best},j} - \mathbf{s}_{j}[k]) + c_2 \xi_2 (\mathbf{g}_{{\rm best}} - \mathbf{s}_{j}[k]),
\end{equation}
\begin{equation}
\mathbf{s}_{j}[k+1] = \mathbf{s}_{j}[k] + \mathbf{v}_{j}[k+1],
\end{equation}
where $\mathbf{s}_{j} = [x_1, y_1, p_1, \dots, x_M, y_M, p_M]$ is the candidate solution vector for all $M$ UAVs. Here, $\omega$ is the inertia weight, $c_1$ and $c_2$ are acceleration coefficients, and $\xi_1, \xi_2$ are random values in $[0,1]$.

For a fair comparison, the initial positions of the particles are derived from the TAKM algorithm, ensuring the baseline begins from the same threat-aware configuration as the proposed MARL scheme. At each time step, the PSO performs an iterative search to maximize the reward function subject to velocity and power constraints, providing a rigorous standard to evaluate the learning-based framework. Specifically, the proposed scheme is compared against two benchmarks: Greedy PSO (GPSO), which optimizes only the UAV trajectories and user associations while transmitting at fixed maximum power, and Optimized PSO (OPSO), which jointly optimizes transmit power in addition to trajectory planning and user association.

\subsection{Computational Complexity and Scalability} 
The computational complexity of our proposed framework is compared with the PSO as a baseline in Table~\ref{tab:performance_complexity}. The complexity of the proposed MATD3 with TAKM clustering and pre-placement is partitioned into two distinct phases: a one-time offline training phase and a real-time online execution phase. The offline training complexity is given by $\mathcal{O}(MKI)$ for clustering, $\mathcal{O}(M^3)$ for UAV matching, $\mathcal{O}\left(Lt_{\max}\left(2M\sum_{i=0}^{L_a} N_a^{(i)}N_a^{(i+1)} + 4\sum_{j=0}^{L_c} N_c^{(j)}N_c^{(j+1)}\right)\right)$ \cite{9826431} for reinforcement learning updates, and $\mathcal{O}(Lt_{\max} MK)$ for calculating user associations during training iterations. The term $I$ is the number of iterations in the TAKM algorithm, $L_a$ and $L_c$ are the number of layers in the actor and critic networks, respectively, and $N_a^{(i)}$ and $N_c^{(j)}$ represent the neuron counts in their respective layers.

Crucially, this intensive computation is performed only once to establish the neural policy. Once the weights are learned and the model is deployed, the critic and target networks are discarded, and only the actor networks are utilized for online operations. Consequently, the online computational complexity scales linearly only with the network architecture, requiring a single forward pass of $\mathcal{O}(\sum N_a^{(i)}N_a^{(i+1)})$ operations. This makes the online execution independent of the total number of  GUs $K$ and eliminates the need for the iterative calculation of the user association matrix $A$ during flight. In contrast, the PSO baseline functions as an iterative search-based optimizer. As shown in Table~\ref{tab:performance_complexity}, its complexity is $\mathcal{O}(MKI) + \mathcal{O}(M^3) + 
\mathcal{O}\!\left(I_{\text{PSO}} N_p (MK + M^2 + MN)t_{\max}\right)$. Unlike the proposed learning-based schemes, the PSO must re-execute its entire iterative search, including the costly $MK$ user-association evaluations for every particle in the swarm, at every time slot to adapt to the dynamic environment.

As illustrated in Table~\ref{tab:performance_complexity}, the proposed MATD3 requires approximately $6.4 \times 10^{10}$ operations for one-time offline training but only $\approx 4.7 \times 10^5$ operations for total swarm online execution per time step. These values are obtained by evaluating the complexity expressions in Table~\ref{tab:performance_complexity} using the simulation parameters from Section VI (e.g., $M = 4$ UAVs, $K = 400$ users, $t_{\text{max}} = 100$ time steps, and the neural network architectures described in Table~\ref{tab:tamarlsim}). Conversely, the PSO baseline requires approximately $1.6 \times 10^{10}$ operations to solve the same problem. Thus, while the proposed framework has a higher initial training cost, it provides a massive reduction in onboard computational burden, enabling real-time autonomous adaptation on resource-constrained UAV hardware.

Notably, the offline complexity of the MADDPG is approximately $5.8 \times 10^{10}$ operations, which is slightly lower than the $6.4 \times 10^{10}$ required by MATD3 because MADDPG utilizes only a single critic per agent rather than a twin-critic architecture. However, during the online phase, both MADDPG and MATD3 exhibit identical computational complexity, as both rely on the forward pass of the same actor network architecture.

Compared to an offline PSO design, the proposed offline MATD3 offers two main advantages. First, MATD3 learns a state-feedback policy that adapts its actions to the current environment. An open-loop PSO plan uses fixed precomputed trajectories and powers and cannot handle deviations from the offline design assumptions (e.g., wind conditions). Such deviations may drive UAVs into threat regions, because the PSO trajectory cannot be corrected once deployed, whereas the proposed MATD3 can adjust its actions online.

Second, if fewer users require service at deployment than assumed during the design stage, MATD3 can adapt its actions to the actual active user set through its state-dependent policy. In contrast, an offline PSO plan cannot update its fixed trajectories and may waste energy flying toward inactive users.

Although the proposed MATD3 framework scales efficiently during decentralized execution, the centralized critic employed under the CTDE paradigm introduces increased training complexity as the number of UAVs grows. Specifically, the critic input dimension scales linearly with the joint observation and action spaces of all UAVs, which may lead to slower convergence during training for very large-scale swarms. Nevertheless, the execution time of the decentralized actors remains independent of the total swarm size since each UAV only requires local observations. To address these centralized training bottlenecks, recent research has explored distributed many-agent perspectives that utilize localized coordination and networked Markov games to decouple learning complexity from the total agent count \cite{10552119}. By focusing on local neighborhood interactions rather than a global state, such distributed architectures provide a scalable roadmap for massive-scale UAV-assisted IoT networks.

\section{Simulation Results}

In our simulation environment, we deploy $M$ UAVs to provide communication services to $K$ GUs distributed non-uniformly across a $4000 \times 4000$ m$^2$ operational area, which contains $L = 2$ distinct threat zones. Specifically, Threat~1 is centered at $(550, 1200)$ m with a radius of $500$ m, while Threat~2 is centered at $(2250, 2500)$ m with a radius of $950$ m. The threat-zone locations were selected to emulate hazardous regions that partially overlap with high-density user areas in practical disaster-response scenarios. The adopted radii (500 m and 950 m) represent threat regions of different spatial scales within the considered operational area, enabling evaluation of the proposed framework under safety constraints. An urban environment is considered. All UAVs operate at a fixed altitude and strictly observe a safety margin around each threat boundary. Unless otherwise specified, the simulation parameters are listed in  Table~\ref{table:SimulationParameters} \cite{10409538, 9826431}.

\begin{table}[t]
\caption{Simulation Parameters}
\label{table:SimulationParameters}
\renewcommand{\arraystretch}{1.3}
\begin{tabular}{|>{\centering\arraybackslash}p{1.5cm}|c|c|c|c|}
\hline
\textbf{Parameter} & $B$ & $K$ & $N$ & $f$ \\
\hline
\textbf{Value} & 10 MHz & 400 & 2 & 1 GHz \\
\hline
\textbf{Parameter} & $\sigma_n^2$ & $P_{\max}$ & $V_{\max}$ & $\delta_t$ \\
\hline
\textbf{Value} & -111 dBm & 33 dBm & 10 m/s & 300 ms \\
\hline
\textbf{Parameter} & $d_{safe}$ & $\zeta_{saf}$ & $d_{\min}$ & $H$ \\
\hline
\textbf{Value} & 10 m & 5 & 10 m & 300 m \\
\hline
\textbf{Parameter} & $a$ & $b$ & $\gamma_{\text{LoS}}$ & $\gamma_{\text{NLoS}}$ \\
\hline
\textbf{Value} & 9.61 & 0.16 & 1 dB & 20 dB \\
\hline
\textbf{Parameter} & $P_0$ & $P_i$ & $U_{\text{tip}}$ & $v_0$ \\
\hline
\textbf{Value} & 9.1827 W & 11.5274 W & 60 m/s & 2.4868 m/s \\
\hline
\textbf{Parameter} & $d_f$ & $\rho$ & $s_r$ & $A_r$ \\
\hline
\textbf{Value} & 0.5017 & 1.205 kg/m$^3$ & 0.0832 & 0.2827 m$^2$ \\
\hline
\end{tabular}
\vspace{-1em}
\end{table}

\subsection{The Comparison of CKM and TAKM Clustering Algorithms}
Figs.~\ref{fig:CKM Clustering} and ~\ref{fig:TAKM Clustering} illustrate the results of applying the CKM and the TAKM algorithms for user clustering, with the distances from the centroids to each threat summarized in Table~\ref{table:distance threats}. It is important to note that the number of UAVs, which is four, is obtained using Algorithm~\ref{alg:adaptive_takm} and represents the minimum number of UAVs required to cover the considered area. The analysis assumes a minimum data rate of $R_{\min} = 0.128$~Mbps. Therefore, $M = 4$ will be used for the rest of the experiments.

\begin{figure}[!t]
\centering
\includegraphics[width=3.5 in]{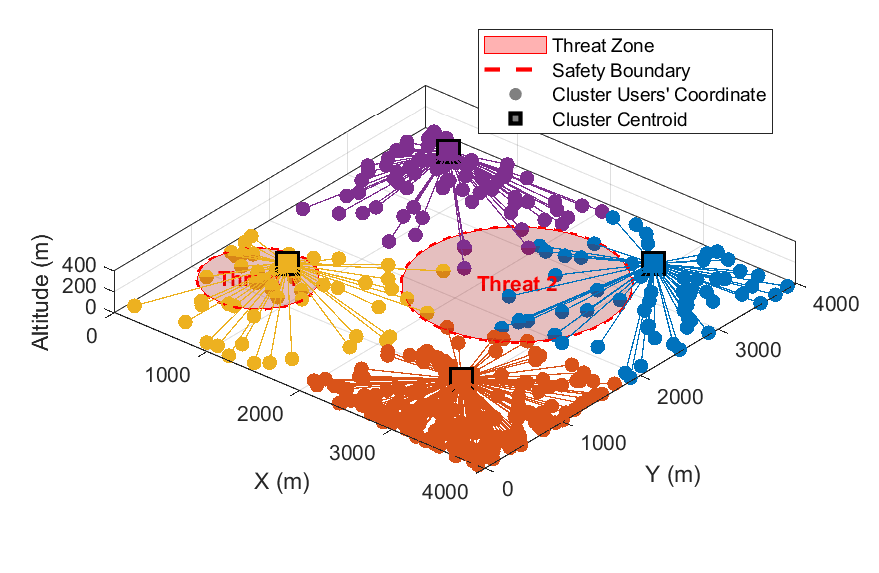}
\vspace{-2.5em}
\caption{ Clustering using the CKM Algorithm.}
\label{fig:CKM Clustering}
\vspace{-1em}
\end{figure}

\begin{figure}[!t]
\centering
\includegraphics[width=3.5 in]{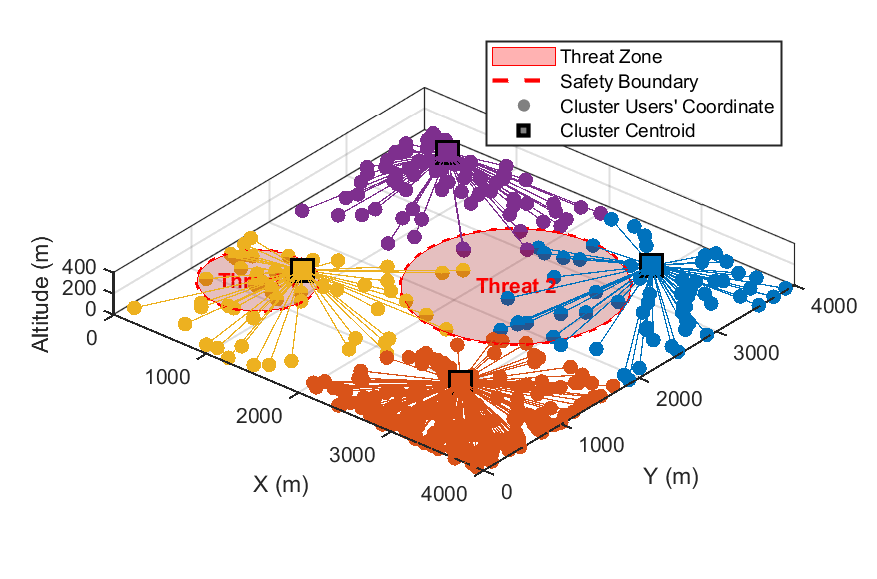}
\vspace{-2.5em}
\caption{ Clustering using the proposed TAKM Algorithm.}
\label{fig:TAKM Clustering}
\vspace{-1em}
\end{figure}

\begin{table}[t]
\centering
\caption{Distance Between CKM/TAKM Cluster Centroids and Threats}
\begin{tabular}{|c|c|c|c|c|}
\hline
\multirow{2}{*}{\textbf{Cluster}} & 
\multicolumn{2}{c|}{\textbf{CKM}} & 
\multicolumn{2}{c|}{\textbf{TAKM}} \\
\cline{2-5}
& \textbf{Threat 1} & \textbf{Threat 2} & \textbf{Threat 1} & \textbf{Threat 2} \\
\hline
Cluster 1 & 3269.50  & 1134.33  & 3274.17  & 1138.50 \\
\hline
Cluster 2 & 2754.85  & 2139.65  & 2766.54  & 2150.54 \\
\hline
Cluster 3 & \textbf{151.31 (UNSAFE)} & 2285.86 & 510 & 1784.66 \\
\hline
Cluster 4 & 2018.21  & 1668.77  & 2240.12  & 1785.94 \\
\hline
\end{tabular}
\label{table:distance threats}
\end{table}

The key advantage of the proposed TAKM method, as visually confirmed in Fig.~\ref{fig:TAKM Clustering}, is its ability to maintain safety guarantees without significantly compromising clustering quality. While the CKM algorithm (Fig.~\ref{fig:CKM Clustering}) minimizes within-cluster variance without regard to constraints, leading to unsafe UAV placements, the proposed TAKM approach incorporates threat avoidance during centroid updates. This ensures compliance with safety requirements while still preserving the convergence properties and computational efficiency of CKM algorithm.

Based on the distance analysis in Table~\ref{table:distance threats}, the comparison between CKM and the proposed TAKM clustering highlights the critical advantage of incorporating threat awareness into the clustering process. Under CKM, Cluster~3 is positioned only 151.31~m from Threat~1, placing the centroid inside the 500~m threat radius and resulting in an unsafe configuration. This unsafe placement reflects a key limitation of CKM in threat-constrained environments, as it forms clusters solely based on geometric proximity. In contrast, the proposed TAKM algorithm successfully relocates all centroids, including Cluster~3, to safe positions, with the TAKM centroid of Cluster~3 lying 510~m from Threat~1, thereby completely outside the hazardous zone. As summarized in Table~\ref{table:Cluster Position}, TAKM achieves this safety-aware repositioning with only slight redistribution in GU assignments (e.g., Cluster~2 changes from 200 to 198 GUs, Cluster~3 from 52 to 55), while maintaining overall cluster balance. These results show that TAKM ensures safe, balanced, and effective UAV clustering while avoiding threat regions.

The matching assignment of UAVs to TAKM centroids is assumed to be optimally solved using the Hungarian algorithm, a well-known method. Since this step incurs a fixed, one-time energy cost and does not affect the comparative evaluation of TAKM and MARL performance, it is not simulated and our experiments focus on the clustering and learning stages only.

\begin{table}[t]
\centering
\caption{Cluster Position and Number of GUs}
\label{table:Cluster Position}
\begin{tabular}{|c|c|c|c|c|}
\hline
\multirow{ 2}{*}{\textbf{Cluster}} & \multicolumn{2}{c|}{\textbf{Cluster Centroid}} & \multicolumn{2}{c|}{\textbf{Number of GUs}} \\ \cline{2-5}
 & \textbf{CKM} & \textbf{TAKM} & \textbf{CKM} & \textbf{TAKM} \\ \hline

Cluster 1 & [3224.35, 3080.82]& [3224.35, 3088.92]  & 65  & 65
\\ \hline
Cluster 2 & [3239.34, 602.82] & [3249.8, 596] & 200  & 198 
\\ \hline
Cluster 3 & [909.04, 1145] & [1060, 1170] & 52  & 55 \\ \hline
Cluster 4 & [739.47, 3209.3] & [727.2, 3433.1] & 83  & 82 \\ \hline

\end{tabular}
\end{table}

\begin{table}[t]
\centering
\caption{MARL Simulation Parameters}
\begin{tabular}{|l|l|}
\hline
\textbf{Parameter} & \textbf{Value} \\ \hline
Number of episodes ($L$) & 400 \\ \hline
Episode duration & 30 s (100 time slots) \\ \hline
Discount factor $\alpha$ & $0.95$ \\ \hline
Soft updated factor $\tau$ & $0.005$ \\ \hline
Actor network & 3 hidden layers, 256 neurons each \\ \hline
Critic network & 3 hidden layers, 256 neurons each \\ \hline
Actor input & UAV 2D coordinates \\ \hline
Actor output & Speed, Flight angle, transmit power \\ \hline
 Critic input & 20 neurons for M=4,  25 neurons \\ & for M=5, 
and 45 neurons for M=9 \\ \hline
Critic output & Q-value of state--action pair \\ \hline
Activation (hidden layers) & ReLU \\ \hline
Actor output activation & Sigmoid \\ \hline
Critic output activation & Linear \\ \hline
Optimizer & Adam \cite{kingma2014adam} \\ \hline
Learning rate $\gamma$ & 0.00001 \\ \hline
Replay memory size & 20,000 \\ \hline
 $E_{max}$& $10^6 b/J$  \\ \hline
\end{tabular}
\label{tab:tamarlsim}
\vspace{-1em}
\end{table}

\subsection{Performance of the Proposed MATD3 with TAKM Clustering and Pre-placement Scheme}

In this section, we evaluate the performance of the proposed MATD3 with TAKM clustering and pre‑placement scheme and compare it with other RL and meta-heuristic benchmark methods. The UAVs are initialized at the centroids obtained from the TAKM algorithm (Table~\ref{table:Cluster Position}). Both the decentralized actor and centralized critic are implemented as fully connected neural networks with three hidden layers. Rectified linear unit (ReLU) activation is used in the hidden layers, with a sigmoid activation at the actor’s output and a linear activation at the critic’s output. Training employs the Adam optimizer \cite{kingma2014adam} and an exploration factor initialized at 0.4, which is multiplied by 0.995 at the beginning of each episode. Detailed training parameters are summarized in Table~\ref{tab:tamarlsim}.

To verify the effectiveness of the reward function design, an ablation study was conducted. Under the baseline configuration ($\zeta_e=1, \zeta_c=1, \zeta_{\mathrm{saf}}=5$), the system achieves an average transmit power of $73$~W while maintaining full coverage. Removing the energy efficiency component ($\zeta_e=0$) resulted in an increase in power consumption to $80$~W, representing a nearly 10\% degradation in efficiency. Similarly, disabling the coverage term ($\zeta_c=0$) led to a reduction in the percentage of served users to 98\%. These results confirm that each reward component is essential for guiding the agents toward an optimized state that balances connectivity and resource conservation. Furthermore, to verify the robustness of the safety weight selection, we performed a sensitivity sweep of $\zeta_{\mathrm{saf}}$ over $\{3, 5, 10, 15\}$. In all cases, the trained agents maintained the required safety distances and exhibited zero entries into prohibited threat regions throughout the testing phase, with negligible variation in EE performance. This confirms that the chosen weighting $\zeta_{\mathrm{saf}}=5$ is robust, while sufficiently large safety weights are essential to enforce collision avoidance and threat evasion.

To demonstrate the unsuitability of CKM clustering in threat-constrained UAV deployment scenarios, Fig. \ref{fig:fig.18} illustrates the safety violations versus training episodes for the MATD3 scheme initialized with CKM clustering and pre-placement. As shown, the safety violation rate remains persistently high, reaching about 70\% even after 400 episodes. This occurs because CKM places UAVs inside threat zones, preventing the MARL agent from learning safe trajectories and resulting in repeated violations even with an advanced algorithm like MATD3. These results highlight a key limitation of conventional clustering in constrained environments: without built-in safety awareness, even sophisticated MARL methods cannot achieve safe UAV deployment. In contrast, the proposed TAKM algorithm explicitly enforces safety during initialization, allowing MARL to focus on optimizing performance rather than escaping unsafe regions. 

Fig.~\ref{fig:fig.5} compares the average reward performance of the proposed MATD3 with TAKM clustering and pre-placement against five benchmark schemes: MADDPG with TAKM clustering and pre-placement, MATD3 with only pre-placement, MADDPG with only pre-placement, MATD3 with random pre-placement, and MATD3 with CKM clustering and pre-placement. Each curve is obtained from one independent training run of 400 episodes, as in [15], and following Algorithm~3. The proposed method achieves the highest final EE among the evaluated schemes and exhibits smooth convergence, reaching a final EE of $4.48 \pm 0.25 \times 10^{6}$~b/J, where the reported mean and standard deviation are calculated over the 100 time slots of the best episode. This represents a notable improvement over MATD3 with only pre-placement ($4.25 \pm 0.26 \times 10^{6}$~b/J) and MATD3 with random pre-placement ($3.66 \times 10^{6}$~b/J). The proposed method also achieves a competitive EE compared to MADDPG with TAKM clustering and pre-placement ($4.41 \pm 0.28 \times 10^{6}$~b/J), while offering faster convergence and more stable learning behavior. The reported standard deviations reflect temporal variations across the time slots of the selected episode, rather than variability across independent training runs. 
While MATD3 with CKM clustering and pre-placement achieves an EE of \(4.45 \times 10^6\)\,b/J, it leads to persistent safety violations, underscoring that threat-aware initialization is essential for real-world deployment  where safety is encouraged via a heavily weighted penalty in the reward function. The results confirm that the TAKM-based initialization is essential for producing safe, threat-aware UAV placements, enabling the reinforcement learning framework to achieve high EE without safety violations.

\begin{figure}[t]
\centering
\includegraphics[width=2.75 in]{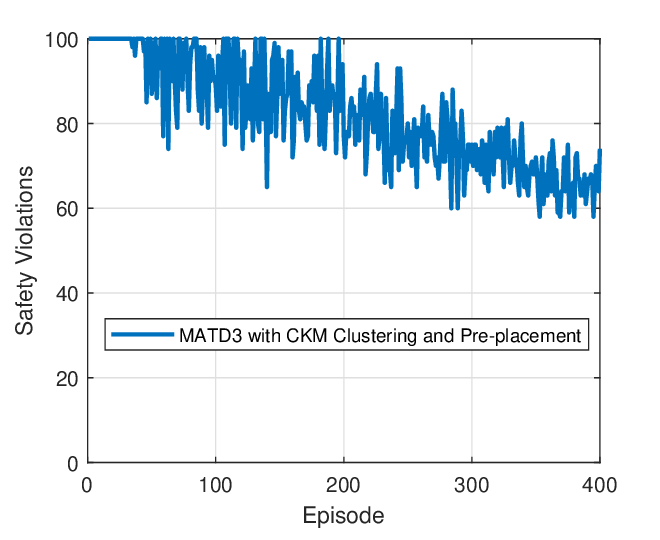}
\vspace{-0.5em}
\caption{Safety violations versus Episodes for MATD3 with CKM clustering and Pre-placement.}
\vspace{-1em}
\label{fig:fig.18}
\end{figure}
\begin{figure}[t]
\centering
\includegraphics[width= 3 in]{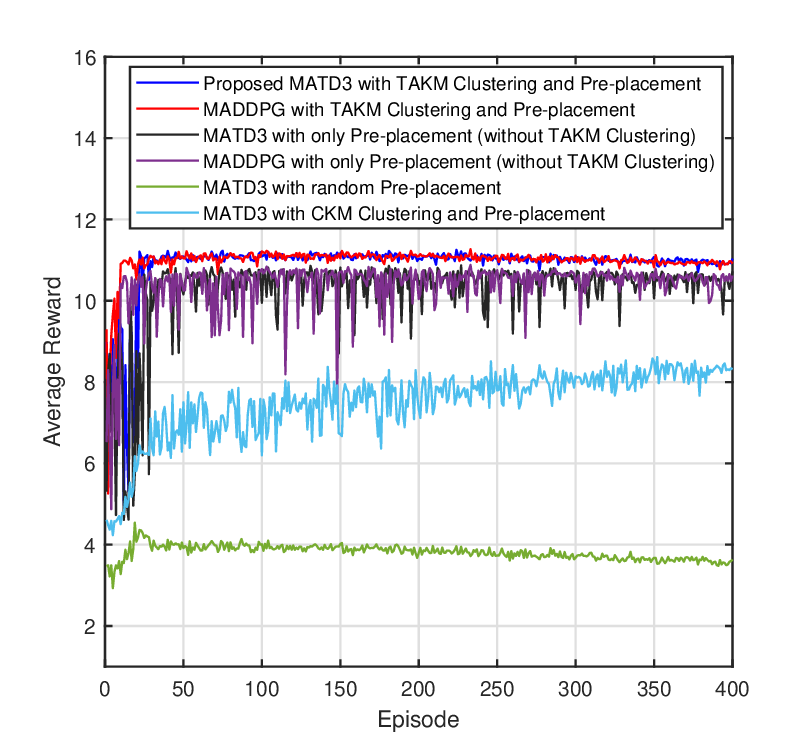}
\vspace{-0.5em}
\caption{The average reward comparison between the proposed scheme and the benchmark schemes.}
\vspace{-1em}
\label{fig:fig.5}
\end{figure}

Fig.~\ref{fig:combined_figures} presents a comprehensive comparison of safety-violation behavior across different schemes, highlighting the clear advantage of the proposed TAKM clustering. The results in Figs.~\ref{fig:combined_figures}(a)  show that MATD3 with TAKM clustering and pre-placement achieves the fastest reduction in violations, approaching nearly zero after roughly 300 episodes. In comparison, MADDPG with TAKM clustering and pre-placement exhibits more safety  violation as shown in Fig.~\ref{fig:combined_figures}(b), even after 300 episodes, reflecting its sensitivity to Q-value overestimation and gradient noise. Figs.~\ref{fig:combined_figures}(c) and~~\ref{fig:combined_figures}(d) illustrate the safety-violation behavior of MATD3 and MADDPG, respectively, when only pre-placement is employed without TAKM clustering. In both cases, frequent safety violations are observed during the training episodes. Although the overall number of violations decreases as training progresses, neither algorithm succeeds in fully eliminating unsafe behavior, even at episode~400, indicating the absence of convergence to a strictly safe policy. MATD3 in Fig.~\ref{fig:combined_figures}(c) exhibits slightly improved stability compared to MADDPG in Fig.~\ref{fig:combined_figures}(d); however, this improvement is insufficient to guarantee safety without threat-aware clustering. These results clearly demonstrate that pre-placement alone is inadequate, and that TAKM clustering is essential to provide structured, threat-aware initial UAV positions that suppress unsafe exploration and enable near-zero safety violations.

\begin{figure*}[t]
    \centering
    \begin{subfigure}[t]{0.24\linewidth}
        \centering
        \includegraphics[width=1\linewidth]{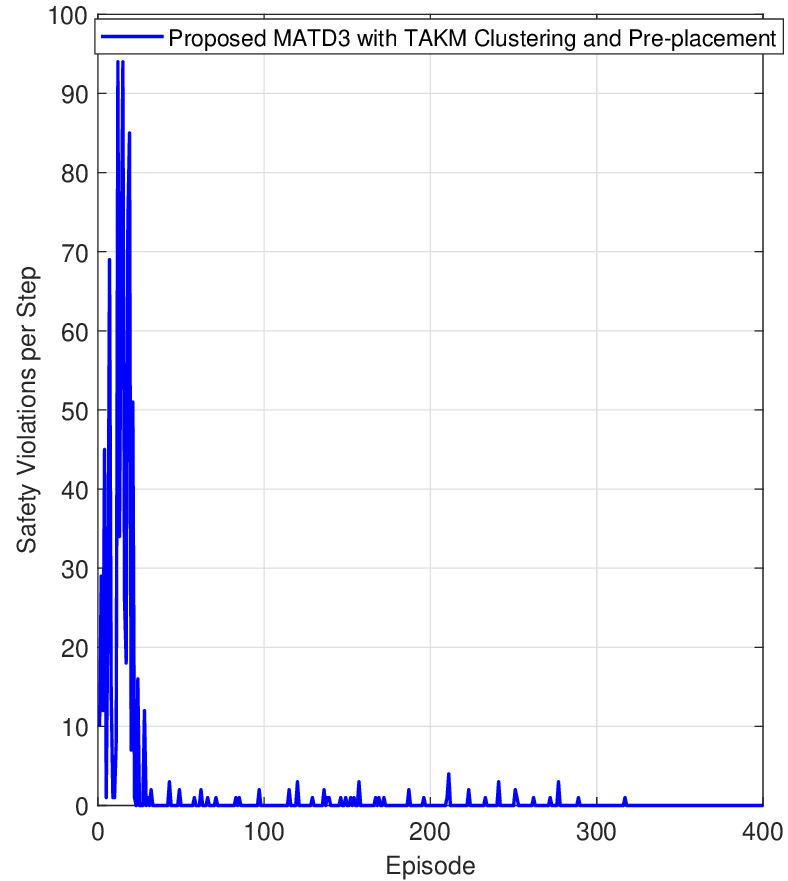}
        \caption{}
        \label{fig:7a}
    \end{subfigure}
    \begin{subfigure}[t]{0.24\linewidth}
        \centering
        \includegraphics[width=1\linewidth]{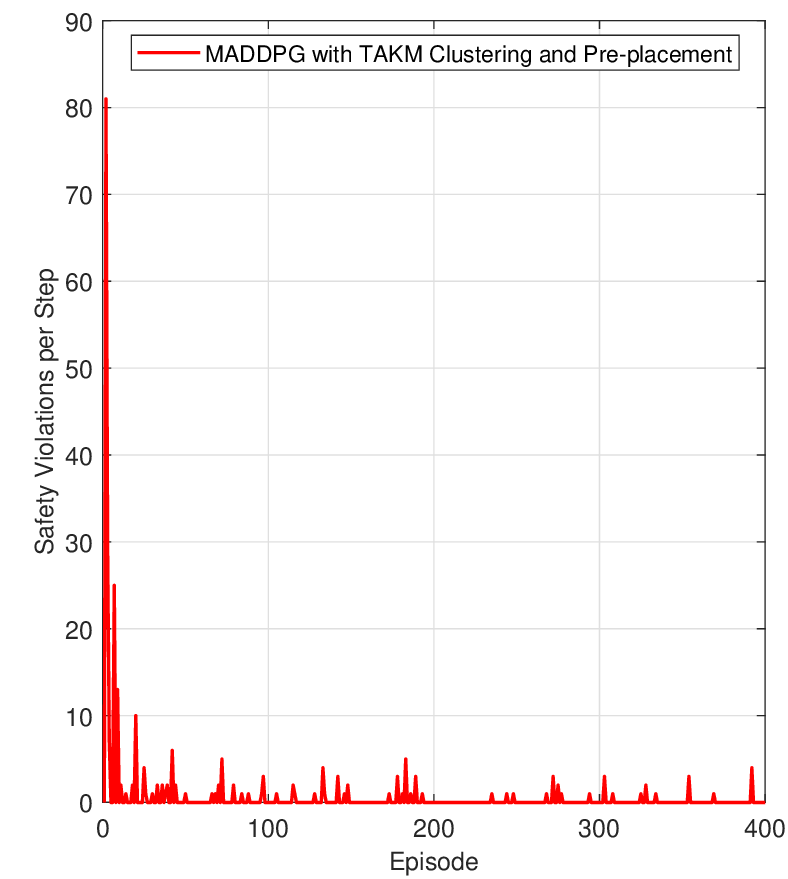}
        \caption{}
        \label{fig:7b}
    \end{subfigure}
    \begin{subfigure}[t]{0.24\linewidth}
        \centering
        \includegraphics[width=1\linewidth]{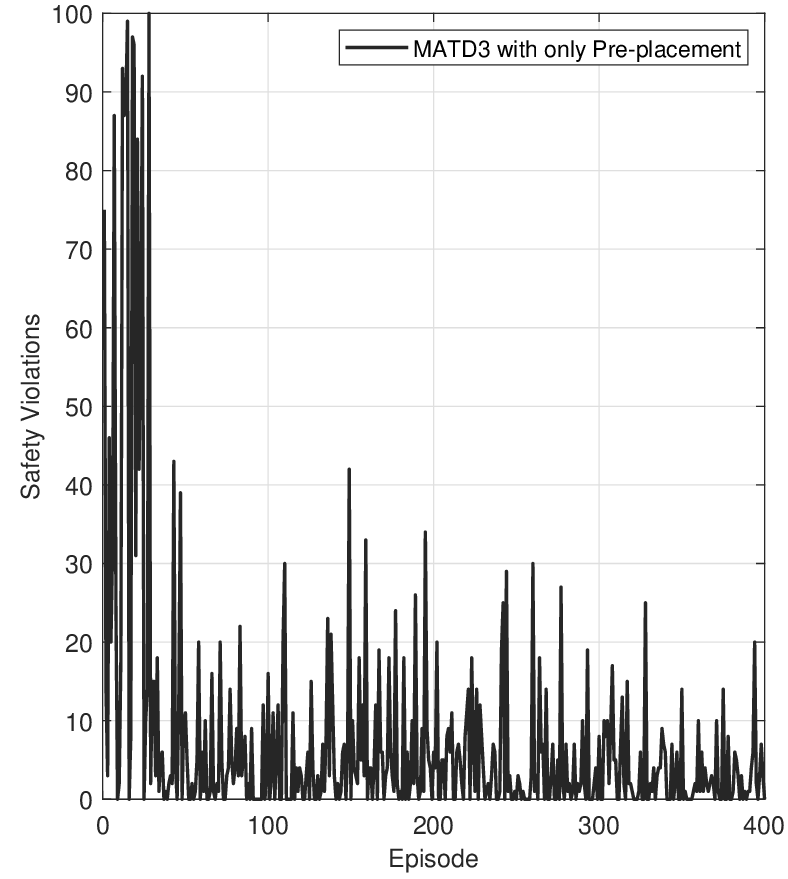}
        \caption{}
        \label{fig:7c}
    \end{subfigure}
    \begin{subfigure}[t]{0.24\linewidth}
        \centering
        \includegraphics[width=1\linewidth]{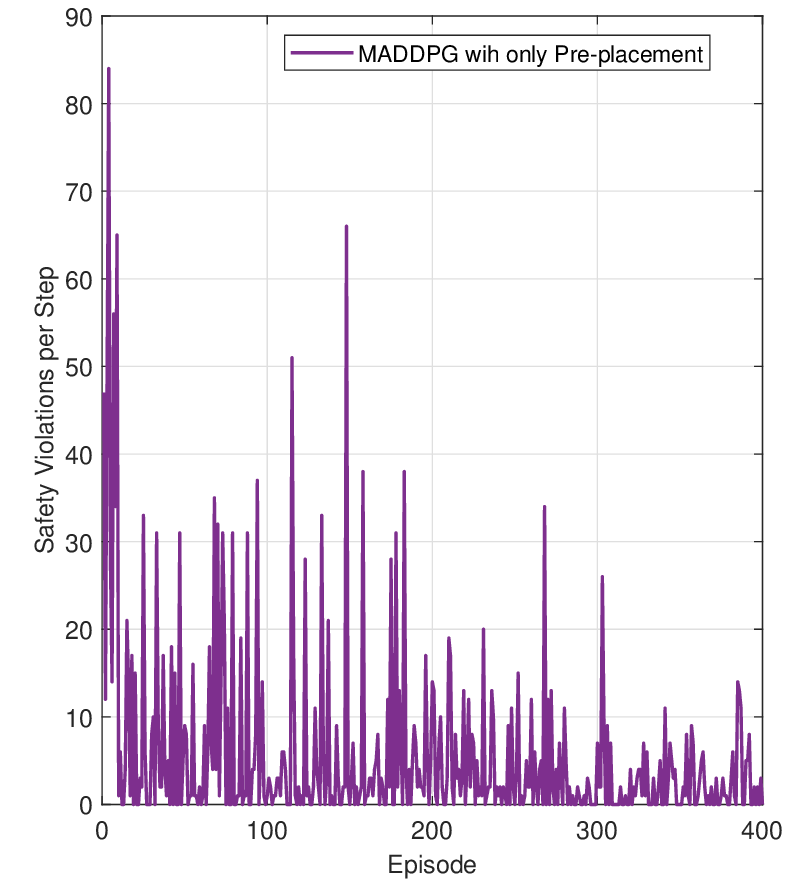}
        \caption{}
        \label{fig:7d}
    \end{subfigure}
    \caption{Safety violations versus Episodes for the proposed scheme and the benchmark schemes.}
    \vspace{-1em}
    \label{fig:combined_figures}
\end{figure*}

\begin{figure}[!t]
\centering
\includegraphics[width=2.75 in]{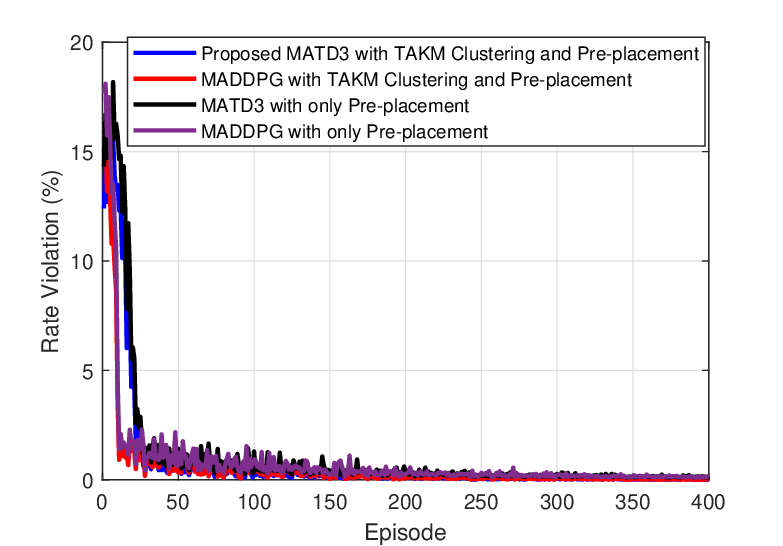}
\caption{Rate violations percentage vs. episodes for the proposed scheme and benchmark schemes.}
\vspace{-1em}
\label{fig:Rateviolation}
\end{figure}

Fig.~\ref{fig:Rateviolation} shows the rate violation percentage versus training episodes for the proposed MATD3-based schemes and the MADDPG benchmark schemes, both with and without TAKM clustering and pre-placement. During the early stages of training, all schemes experience a high rate violation percentage due to random exploration and untrained policies. However, a rapid decline in violations is observed within the initial episodes. Notably, the MADDPG-based schemes exhibit a faster initial reduction in rate violations compared to the proposed MATD3 schemes. Nevertheless, the proposed MATD3 with TAKM clustering and pre-placement demonstrates stable convergence, similar to MADDPG with TAKM clustering and pre-placement, and achieves comparable violation levels in later episodes. After approximately 150–200 episodes, the rate violation percentages of all schemes converge toward zero, indicating that all users are successfully served. 

\begin{figure}[!t]
\centering
\includegraphics[width=2.75 in]{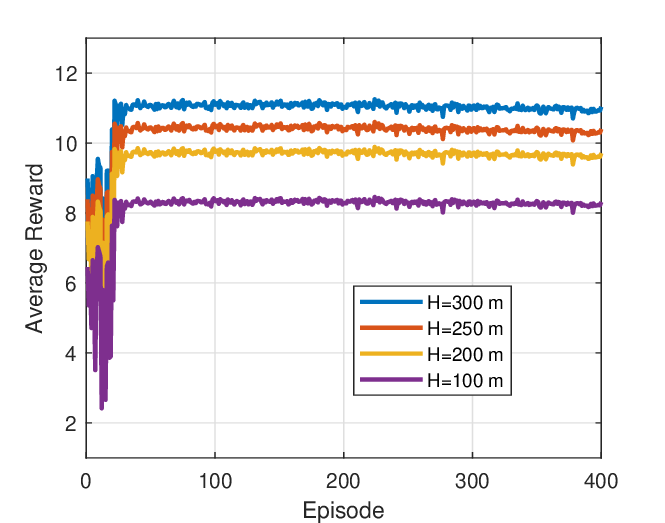}
\caption{The average reward performance with different altitudes for the proposed scheme.}
\vspace{-1em}
\label{fig:different altitudes}
\end{figure}
\begin{figure}[!t]
\centering
\includegraphics[width=2.75 in]{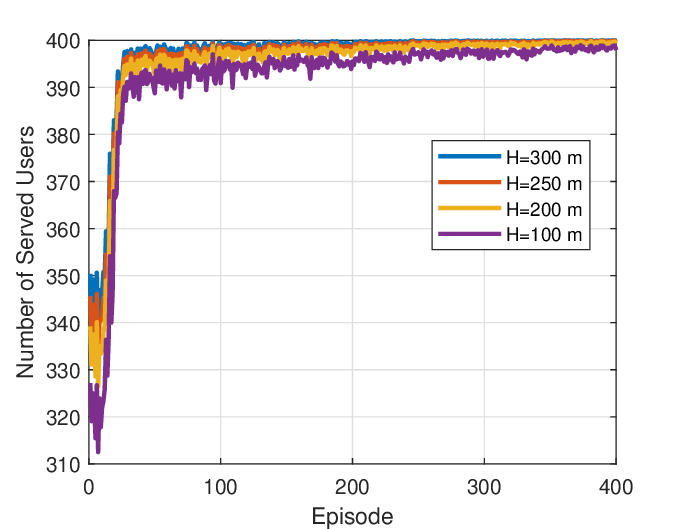}
\caption{Number of served users at different altitudes for the proposed scheme.}
\vspace{-1em}
\label{fig:served users}
\end{figure}
\begin{figure}[!t]
\centering
\includegraphics[width=2.75 in]{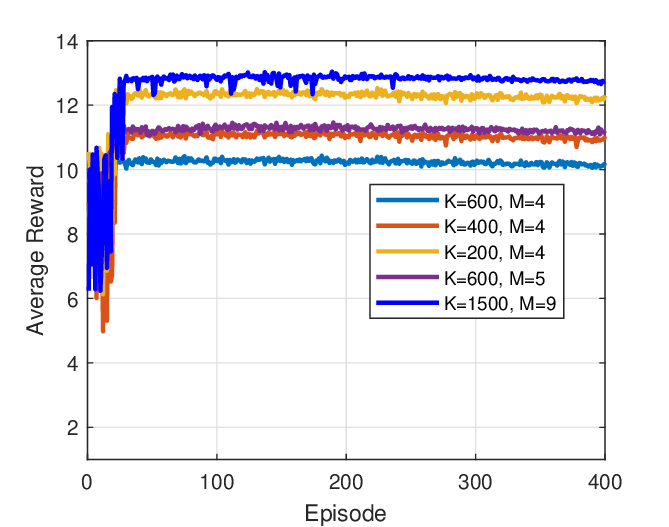}
\caption{The average reward performance with different number of users for the proposed scheme.}
\vspace{-1em}
\label{fig:different number of users}
\end{figure}

Figs. ~\ref{fig:different altitudes} and ~\ref{fig:served users} collectively emphasize the critical role of UAV altitude in optimizing overall system performance.    The considered UAV altitudes (100--300 m) represent practical low-to-moderate altitude operating ranges commonly adopted in urban UAV communication studies. These values enable evaluation of the tradeoff between blockage reduction and LoS improvement at higher altitudes versus the increased propagation distance and energy consumption associated with excessive elevation. In particular, the choice of the maximum altitude $H=300$ m is grounded in the system configuration adopted in \cite{9826431}. Fig. ~\ref{fig:different altitudes} shows that an altitude of 300 $m$ delivers the highest average reward, substantially outperforming 250 $m$, 200 $m$ and 100 $m$. This superior performance is primarily due to improved channel conditions at higher altitudes: as UAVs gain height, the probability of establishing LoS links with GUs increases according to the probabilistic channel model. Higher LoS probability reduces path loss, enhances signal quality, and allows for higher data rates with more efficient power consumption.

Fig.  ~\ref{fig:served users} provides complementary insights by highlighting the effect of altitude on user coverage. At 300 $m$, the UAVs can consistently serve all 400 GUs, whereas the other altitudes cannot fully cover all users. This demonstrates that the superior reward at 300 $m$ is not only a result of better channel conditions but also due to meeting the system’s primary requirement of universal service. Together, these figures illustrate the same altitude that ensures full coverage simultaneously maximizes reward, channel rate, and EE. In threat-constrained environments with non-uniform user distribution, higher altitudes therefore offer the dual benefit of achieving complete service coverage and optimal communication performance, making them the most effective choice for UAV deployment. Given these results, an altitude of 300 $m$ is selected for the proposed scheme in this paper. 

Fig.~\ref{fig:different number of users} illustrates the average reward obtained under different numbers of users and shows that the reward decreases as the number of users grows. This behavior is due to the system's operational constraints: as the number of users increases, the UAV transmit power is allocated among more GUs. Given the fixed maximum transmit power per UAV, the power allocated to each user decreases. Moreover, each user is allocated less bandwidth, which reduces the average channel capacity and consequently lowers the reward. These results demonstrate that the proposed learning-based approach can adapt to scenarios with varying user densities while maintaining efficient operation. It is also noted that the coverage rate remains at \(100\%\) for 200 and 400 users, but decreases to \(95\%\) for 600 users with \(M=4\). However, when the number of UAVs is increased to \(M=5\) (determined via Algorithm~\ref{alg:adaptive_takm} for \(K=600\)), the coverage is restored to \(100\%\) and the average reward increases compared to the case with \(M=4\). This improvement occurs because with more UAVs, each UAV serves fewer users, leading to higher transmit power per user, better channel conditions, and more efficient bandwidth allocation, thereby enhancing overall system EE. 

To further evaluate scalability, we consider a larger scenario with $K=1500$ users and the result is shown in Fig.~\ref{fig:different number of users}. Algorithm~\ref{alg:adaptive_takm} first determines that $M=9$ UAVs are required to satisfy the network demand, after which MATD3 optimizes their deployment and resource allocation. For the $M=9$ scenario, the critic input dimension was scaled to $45$. This scaling is consistent with the centralized critic architecture described in Section~V-B.} Under this configuration, the proposed framework achieves $100\%$ user coverage with zero safety violations and zero rate violations. Despite the larger swarm size, the CTDE-based MATD3 maintains stable convergence, indicating that the centralized critic remains computationally manageable for the considered deployment scale.

The scalability results in Fig.~\ref{fig:different number of users} demonstrate that the proposed framework scales effectively by first determining the required number of UAVs and their initial safe positions via Algorithm~\ref{alg:adaptive_takm}, and then optimizing their trajectories, transmit powers, and user associations with MATD3. Consequently, the framework preserves its safety, coverage, and QoS guarantees as the network size increases. 

\begin{table}[t]
\centering
\caption{Comparison of User Distribution.}
\label{tab:Distribution_comparison}
\begin{tabular}{|c|c|c|c|c|c|}
\hline
\textbf{Method} & 
\textbf{Cluster 1} & 
\textbf{Cluster 2} & 
\textbf{Cluster 3} & 
\textbf{Cluster 4} & 
\textbf{SDNU} \\
\hline
CKM & 65 & 200 & 52 & 83 & 58.78 \\
\hline
TAKM & 65 & 198 & 55 & 82 & 57.40 \\
\hline
MATD3 & 64 & 183 & 67 & 86 & 48.66 \\
\hline
\end{tabular}
\vspace{-1em}
\end{table}

\begin{figure}[!t]
\centering
\includegraphics[width=3 in]{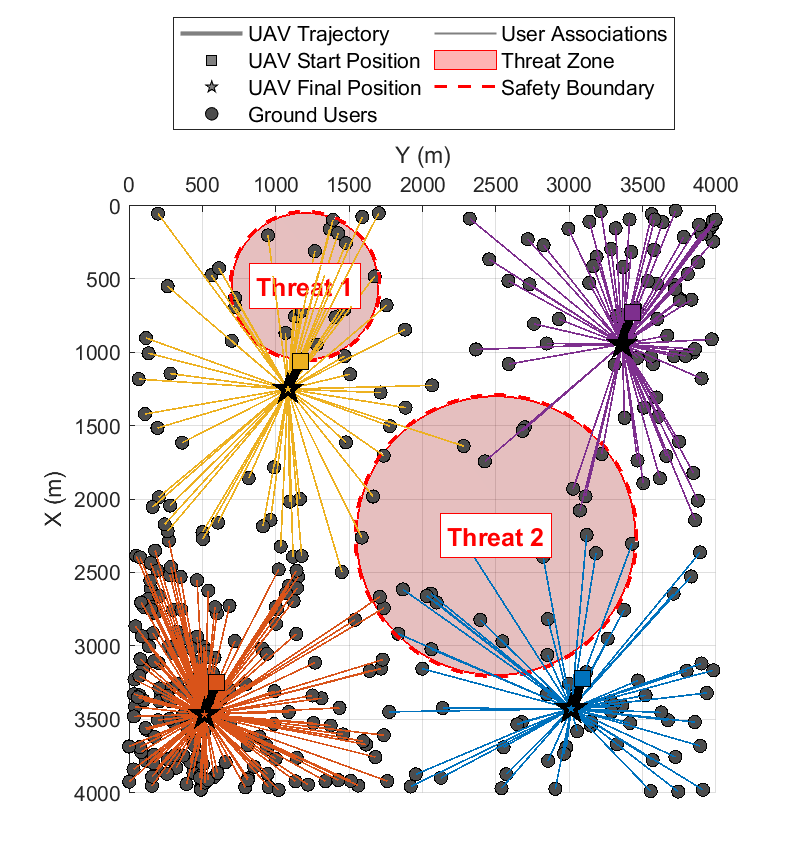} 
\vspace{-0.5em}
\caption{The trajectories of UAVs for the proposed scheme.}
\vspace{-1em}
\label{fig:MATD3 trajectory}
\end{figure}

Fig. ~\ref{fig:MATD3 trajectory} illustrates the UAV trajectories generated by the proposed MATD3 scheme with TAKM clustering and pre-placement. The initial UAV positions correspond to the TAKM centroids in Table ~\ref{table:Cluster Position}, where the four clusters contain 65, 198, 55, and 82 GUs, respectively. These values represent the initial user grouping using TAKM. However, as the UAVs follow the learned trajectories, they gradually reposition themselves toward more energy-efficient locations, which reshapes the effective coverage regions and results in new user grouping.  This dynamic re-clustering is reflected in the final user distributions, where the UAVs ultimately serve 64, 183, 67, and 86 GUs, and their final horizontal positions are (3429.8, 3013.3)\,m, (3467.4, 514.6)\,m, (1251.3, 1082.9)\,m, and (945.2, 3360.8)\,m, respectively. The changes in user assignment demonstrate that the proposed method does not rely on fixed clustering; instead, the UAVs adapt their service regions to maximize EE while satisfying LoS and safety constraints. UAV~2, initially responsible for 198 GUs, ends up serving 187 users by shifting slightly toward the densest region, whereas UAVs 1, 3, and 4 experience moderate adjustments as their trajectories refine the cluster shapes. 

\begin{figure}[!t]
\centering
\includegraphics[width=3 in]{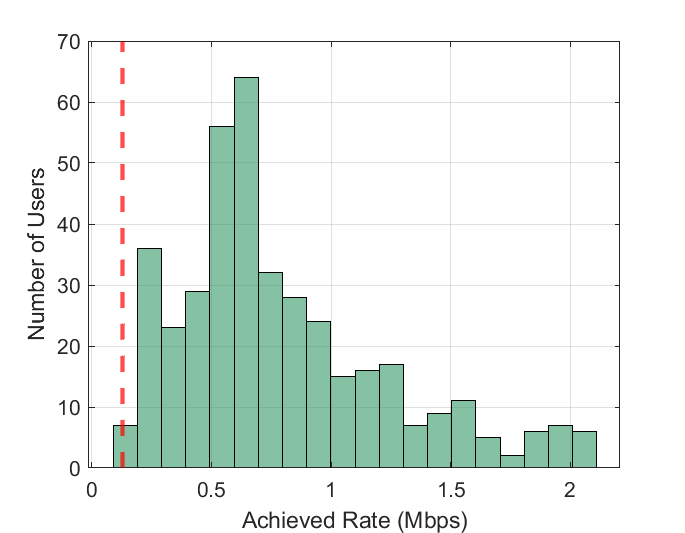}
\caption{Distribution of user rates at final UAV positions.}
\vspace{-1em}
\label{fig:User_rate_final_UAV_Positions}
\end{figure}

The standard deviation of the number of users (SDNU) across clusters is presented in Table ~\ref{tab:Distribution_comparison} for CKM, TAKM, and MATD3. The results show that users are distributed more evenly by the MATD3 framework, with the SDNU decreasing from 58.78 (CKM) to 57.40 (TAKM) and finally to 48.66 (MATD3). This progressive improvement of 10.12\% from CKM to MATD3 occurs because the MATD3 framework dynamically adjusts user associations during trajectory optimization. Compared to static clustering, MATD3 further considers  channel conditions and UAV positioning, resulting in a more balanced workload distribution.

Fig.~\ref{fig:User_rate_final_UAV_Positions} illustrates the distribution of user rates achieved at the final point of the UAV trajectories under the proposed MATD3 with TAKM clustering and pre-placement scheme. The red line in the figure indicates the minimum rate $R_{\min}$. Fig.~\ref{fig:User_rate_final_UAV_Positions} shows that the distribution exhibits a pronounced right-skew, concentrated above \(R_{\min}\) and extending toward higher rate values, indicating that most users enjoy robust channel conditions while a small subset benefits from particularly favorable links. A per-UAV breakdown further reveals: UAV 1 serves 64 users with an average rate of 1.31 Mbps, UAV 2 serves 183 users at average rate of 0.45 Mbps, UAV 3 serves 67 users at average rate of 1.06 Mbps, and UAV 4 serves 86 users at average rate of 0.96 Mbps, illustrating how the framework dynamically balances load while maintaining high per-user throughput. 

\begin{figure}[!t]
\centering
\includegraphics[width=3 in]{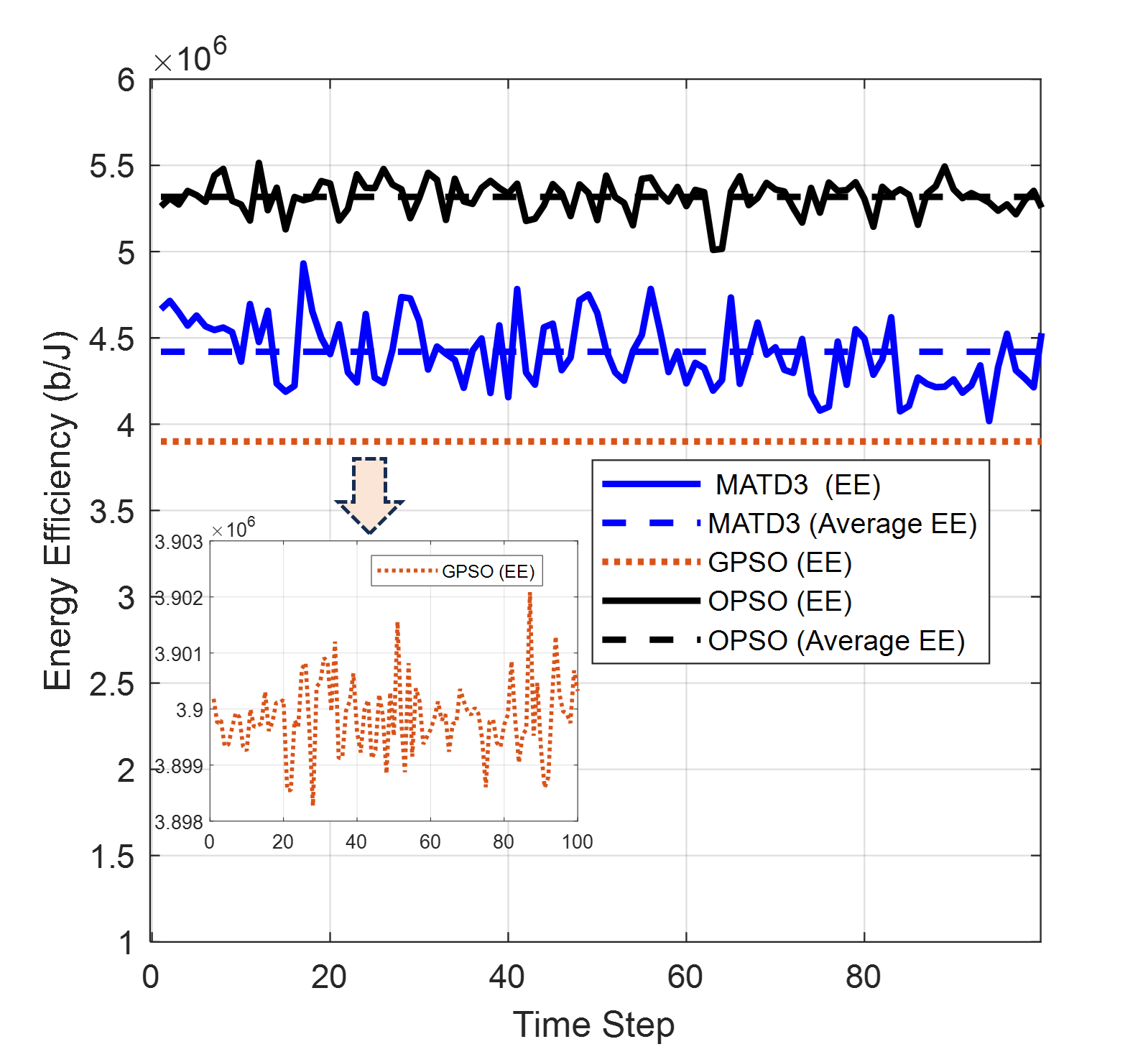}
\caption{Energy efficiency versus time steps for the final trajectories in the proposed and PSO schemes.}
\vspace{-1em}
\label{fig:EE-MATD3-PSO}
\end{figure}

Fig.~\ref{fig:EE-MATD3-PSO} compares the EE of the proposed MATD3 scheme with that of the GPSO and OPSO benchmarks. GPSO denotes a baseline in which each UAV transmits at its maximum power, whereas OPSO
jointly optimizes the UAV trajectories and transmit powers at every time step.

 For the PSO benchmarks, the numbers of particles and iterations were selected before the final performance comparison through a predefined parameter search. Specifically, we evaluated
$N_p\in\{100,150,250,300\}$ and
$I_{\mathrm{PSO}}\in\{100,200,300,400,500\}$, while fixing the inertia weight
to $\omega=0.7$ and the acceleration coefficients to $c_1=c_2=1.5$. The
configuration $N_p=250$ and $I_{\mathrm{PSO}}=400$ was selected, as further increasing the search budget yielded negligible EE improvement. These parameters were then fixed for all benchmark comparisons and were not adjusted based on MATD3 results. Under this setting, OPSO achieved an EE of $5.30\times10^{6}$~b/J, whereas the GPSO baseline achieved $3.90\times10^{6}$~b/J. The obtained results from one training run indicate that MATD3 achieves a higher EE than the GPSO baseline. This performance difference arises because MATD3 functions as an intelligent learner: its trained actor network predicts velocity and power settings based on a learned policy, whereas GPSO follows a rigid, precomputed path. Although OPSO attains a higher EE, its computational demands make it impractical for onboard UAV implementation, as explained in Section~V-D. In contrast, MATD3 combines low computational complexity, requiring only a single forward pass through the neural network during online execution, with competitive performance compared with traditional heuristic approaches.

The validation results presented in Table \ref{tab:robustness} demonstrate the robustness and reproducibility of the proposed MATD3 framework. In this evaluation setup, the random number generator (rng) seed determines the spatial distribution of GUs within the simulation area. To assess training stability, the proposed method is evaluated across multiple random seeds; the results show only minor performance variation, indicating that the learned policy consistently converges to near-optimal solutions and is not sensitive to stochastic initialization. Initially, the policy was evaluated on the same user distribution used during training (rng 1). In this case, the learned policy achieves $4.24 \times 10^6$ b/J, which corresponds to $94.6\%$ of the peak training performance ($4.48 \times 10^6$ b/J). This minor gap is expected when removing exploration noise during the testing phase and confirms that the policy is stable rather than overfitted to exploratory behaviors.

To further verify generalization, the framework was tested across diverse spatial realizations without retraining or fine-tuning. As shown in Table \ref{tab:robustness}, while the peak training performance reached $4.48 \times 10^6$ b/J, the testing trials yielded a high average EE of approximately $4.13 \times 10^6$ b/J. By retaining over $92\%$ of the training performance across all unseen distributions, these results demonstrate that the MATD3 framework, integrated with TAKM clustering, learns generalized decision-making strategies rather than merely memorizing specific user coordinates. The training peak of $4.48 \times 10^{6}$ b/J in Table \ref{tab:robustness} represents the maximum EE achieved over a single episode during training, whereas the testing values represent averages over multiple independent test runs with exploration noise removed.

\begin{table}[t]
\centering
\caption{Reproducibility and Robustness Evaluation of the MATD3 Framework}
\label{tab:robustness}
\setlength{\tabcolsep}{10pt}
\begin{tabular}{|l|c|c|}
\hline
\textbf{Scenario} & \textbf{User Distribution} & \textbf{EE (b/J)} \\ \hline
Training  & rng (1) & $4.48 \times 10^6$ \\ \hline
Testing (Same) & rng (1) & $4.24 \times 10^6$ \\ \hline
Testing (Different) & rng (4) & $4.10 \times 10^6$ \\ \hline
Testing (Different) & rng (10) & $4.06 \times 10^6$ \\ \hline
Testing (Different) & rng (11) & $4.12 \times 10^6$ \\ \hline
Testing (Different) & rng (14) & $4.22 \times 10^6$ \\ \hline
Testing (Different) & rng (17) & $4.14 \times 10^6$ \\ \hline
Testing (Different) & rng (22) & $4.14 \times 10^6$ \\ \hline
Testing (Different) & rng (27) & $4.07 \times 10^6$ \\ \hline
Testing (Different) & rng (30) & $4.14 \times 10^6$ \\ \hline
\end{tabular}
\vspace{-1em}
\end{table}

\begin{figure}[!t]
\centering
\includegraphics[width=3.35 in]{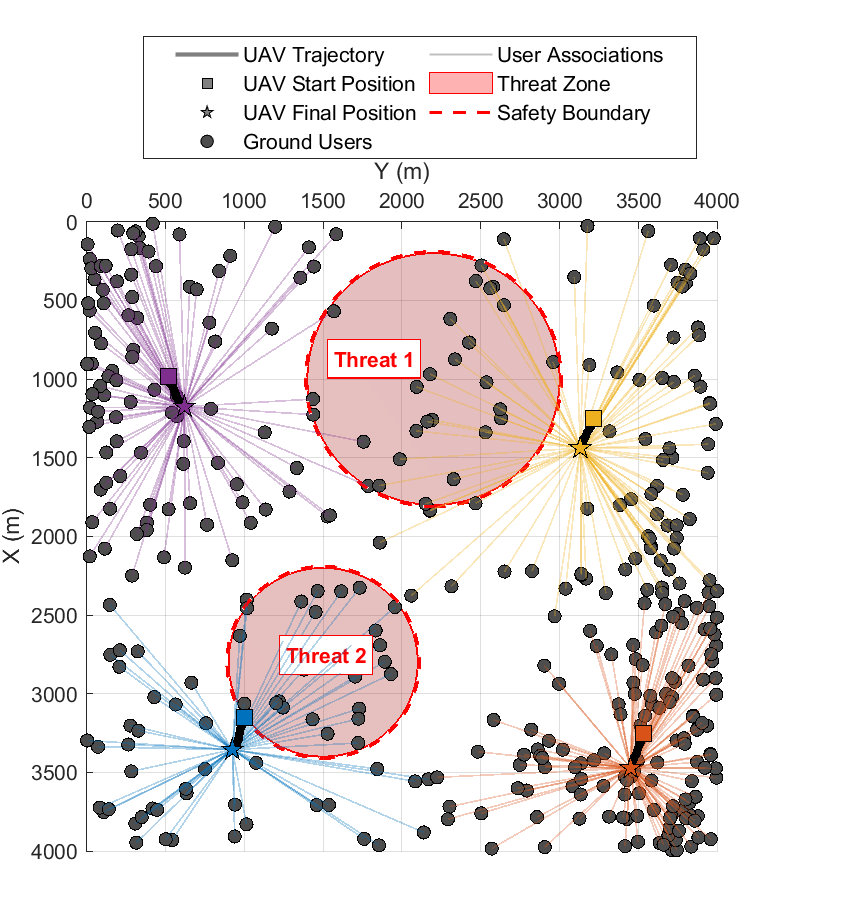}
\vspace{-0.5em}

\caption{UAV trajectories under a new threat configuration with different zone centers and radii.}
\vspace{-1em}
\label{fig:Threat_config_2}
\end{figure}

To assess the safety guarantees of the proposed framework, we tested it under a new threat configuration with different placements and sizes. Specifically, the original two threat zones were replaced by new regions centered at $(1000, 2200)$~m and $(2800, 1500)$~m with radii of $800$~m and $600$~m, respectively, while all other simulation parameters remained unchanged. Under this new configuration, the proposed TAKM algorithm successfully places all UAVs in safe positions, as shown in Fig.~\ref{fig:Threat_config_2}. The results show that the proposed TAKM-MATD3 framework achieved an energy efficiency of $4.13 \times 10^{6}$~b/J, maintained $100\%$ user coverage, and exhibited zero safety violations, with safe trajectories. These results confirm that the proposed threat-aware framework generalizes well to different threat geometries while consistently satisfying the safety constraints.

\begin{figure}[!t]
\centering
\includegraphics[width=3.35 in]{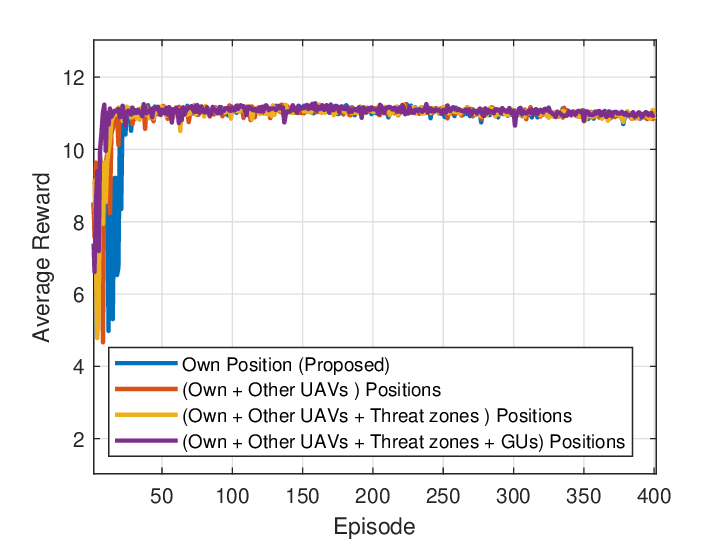}
\vspace{-0.5em}
\caption{Average reward versus episode for the proposed scheme at different observation scenarios.}
\vspace{-1em}
\label{fig:Observations}
\end{figure}

To study the robustness of the proposed restricted observation design, we conducted a nested ablation study in which the observation was progressively extended to include other UAVs' positions, threat-zone locations, and user positions. As shown in Fig.~\ref{fig:Observations}, all four observation configurations converge to nearly identical average reward. During the earlier training episodes, configurations with additional observed information exhibit marginally faster and smoother convergence, while the restricted observation shows larger transient oscillations before stabilizing. However, since this difference is confined to the early training phase and vanishes at convergence, all four configurations maintain zero safety violations and $100\%$ user coverage throughout. Consequently, the restricted observation in the proposed design is sufficient for the fixed scenario considered, and is chosen to reduce communication overhead and execution complexity.

\section{Conclusion}
In this paper, we proposed a threat-aware, energy-efficient MARL framework for deploying multiple UAVs in constrained and hazardous environments. First, we introduced a safety-critical TAKM clustering algorithm (Algorithm~\ref{alg:threat_aware_kmeans}), which embeds safety constraints directly into the clustering process by ensuring that all UAV placements remain outside threat zones with a prescribed safety margin. Second, we integrated TAKM with a systematic procedure (Algorithm~\ref{alg:adaptive_takm}) to determine the minimum number of UAVs and the initial safe UAV positions required to satisfy the design constraints such as  rate, and safety. Third, building upon this safe and efficient initial configuration, we proposed a threat-aware MATD3 framework (Algorithm~\ref{alg:MATD3}) to jointly optimize UAV trajectories, transmit power allocation, and user associations, learning cooperative policies that maximize long-term EE while continuously respecting all operational constraints. 

Extensive simulation results demonstrated that the proposed MATD3 framework with TAKM-based clustering and pre-deployment initialization outperforms benchmark schemes by achieving higher EE, faster convergence, and zero safety violations. The proposed approach surpassed GPSO and approached OPSO performance while requiring significantly lower online deployment computational complexity. The results further demonstrated robust generalization to unseen user distributions, retaining more than 92\% of the training performance without retraining while maintaining zero safety violations under different threat configurations. These findings demonstrate that integrating TAKM with MATD3 provides a practical and scalable solution for safe and energy-efficient UAV deployment in threat-constrained environments. The results also highlight that learning-based algorithms without TAKM clustering are insufficient to guarantee safety, underscoring the critical role of TAKM in safety-aware UAV deployment.

 While the framework was validated in simulation using well-established system models, real flight effects such as wind, GPS drift, and latency were not considered. Consistent with existing research, hardware validation on Pixhawk-based multi-UAV platforms under real-world disturbances is therefore left for future work. Future work may also extend this framework to more complex scenarios such as dynamic hazards, dynamic users, while integrating PER to enhance learning efficiency.

\bibliography{Man_Ref}
\bibliographystyle{ieeetr}

\end{document}